# Blunt-end driven re-entrant ordering in quasi two-dimensional dispersions of spherical DNA brushes

*Ivany Romero-Sanchez,*[1,2]‡ *Ilian Pihlajamaa,*[3,4]‡ *Natasa Adžić,*[3] *Laura E. Castellano,*[2] *Emmanuel Stiakakis,*[5] *Christos N. Likos,*[3]* *Marco Laurati*[1]*

[1]Dipartimento di Chimica & CSGI, Università di Firenze, 50019 Sesto Fiorentino, Italy

[2]División de Ciencias e Ingenierías, Universidad de Guanajuato, 37150 León, Mexico

[3]Faculty of Physics, University of Vienna, Bolzmanngasse 5, A-1090 Vienna, Austria

[4]Eindhoven University of Technology, Department of Applied Physics, Soft Matter and Biological Physics, Postbus 513, NL-5600 MB Eindhoven, The Netherlands

[5]Biomacromolecular Systems and Processes, Institute of Biological Information Processing (IBI-4), 4 Forschungszentrum Jülich, D-52425 Jülich, Germany

**KEYWORDS** Blunt-ends, Colloids, DNA, Polyelectrolytes, Order Transitions

**ABSTRACT**

We investigate the effects of crowding on the conformations and assembly of confined, highly charged, and thick polyelectrolyte brushes in the osmotic regime. Particle tracking experiments on increasingly dense suspensions of colloids coated with ultra-long double stranded DNA (dsDNA) fragments reveal non-monotonic particle shrinking, aggregation and re-entrant ordering. Theory and simulations show that shrinking is induced by the osmotic pressure exerted by the counterions absorbed in neighbor brushes, while aggregation and re-entrant ordering are the effect of a short-range attraction competing with the electrostatic repulsion. Blunt-end interactions between dsDNA fragments of neighboring brushes are responsible for the attraction and can be tuned by inducing free-end backfolding through the addition of monovalent salt. Our results show that base stacking is a mode parallel to hybridization to steer colloidal assembly, in which attractions can be fine-tuned through salinity and, potentially, grafting density and temperature.

Polyelectrolyte brushes, which consist of charged polymer chains grafted to a planar or curved surface, have found key applications due to their unique properties[1,2]. Among them, the ability to prevent protein absorption, called anti-biofouling effect[3,4] and exploited in biomedical applications and tissue engineering[5]. Conversely, at low ionic strength the opposite effect is observed, i.e. strong protein absorption[6]. The two effects can be combined to obtain a protein delivery mechanism[7]. In addition, spherical polyelectrolyte brushes (SPBs) can be used as nanoreactors for the synthesis of metallic nanoparticles with strong catalytic activity[8]. Polyelectrolyte brushes also present an exceptionally low mutual friction[9,10], which arises from the huge osmotic pressure generated by counterions absorbed within the brush in low ionic strength environments, and can be controlled through the addition of multivalent ions[11,12]. The ultra-low friction of polyelectrolyte brushes is essential in biolubrication, for example for the correct operation of synovial joints[13] and the production of coatings for medical implants[14]. In the context of polyelectrolyte-mediated lubrication effects, the interactions between contacting brushes and the resulting polymer conformations play a fundamental role: compression or interpenetration can change friction by orders of magnitude[11]. Such interactions and conformations are decisively influenced by the packing of brushes, in particular in crowded conditions, an aspect which is especially relevant in biological systems. These deformations under crowding, which constitute one of the main issues of the present work, have not been fully explored to date.

DNA is a highly charged polyelectrolyte whose properties have been used to develop nanotechnologies such as electrochemical sensors[15], field-effect transistors[16] and smart surfaces[17]. Being highly customizable with molecular precision, DNA is thus the ideal building block for the systematic investigation of polyelectrolyte brush interactions in crowded systems. In addition, the specificity of DNA interactions can be exploited to precisely control assembly. Complex DNA structures can be assembled through Watson-Crick base pairing of sticky ends[18], the programmable folding of long single strands (DNA Origami and "brick" assembly)[19] or supramolecular interactions[20], including blunt-end base stacking[21–23]. DNA can be also used to direct the assembly of colloidal micro- and nanoparticles, which can confer to materials desired optical, electrical or mechanical properties[24,25]. For colloidal systems base-pairing interactions between sticky ends have been the main tool used to construct desired structural arrangements[26,27]. Blunt-end interactions, while scarcely considered[28], provide a parallel route to colloidal assembly that exploits base stacking instead of base-pairing. The effective interaction between a collection of blunt-end of dsDNA fragments grafted onto two facing colloidal particles can be fine-tuned through ionic strength and ion type, and, potentially, grafting density and temperature.

In this work, we demonstrate that the combination of strong osmotic forces from the neighboring SPBs and the blunt-end DNA attractions leads to novel, re-entrant ordering phenomena and to the emergence of string-like patterns in concentrated DNA-brush solutions. To this end, we combined microscopy experiments, theory and simulations to systematically investigate the interactions and correlations between thick, densely packed SPBs in quasi-2D confinement. In these experiments the structural organization and the dynamics of thick spherical dsDNA brushes grafted onto latex beads were determined. Two main effects of the brush-brush interactions were found: a progressive reduction of the inter-particle effective interaction range (distance of closest approach) with increasing packing and a complex, unusual aggregation behavior, with a re-entrant ordering as a function of packing fraction. By developing a detailed microscopic model of the effective interactions, which incorporates electrostatic, entropic, and osmotic free energy contributions, as well as the concentration-dependent blunt-end attractions, we determined the conformations of single and contacting SPBs. Moreover, we established that the decrease of the inter-particle distance is associated with a size reduction that origins from the pressure exerted on a brush by the absorbed, non-condensed counterions of neighbor brushes. This mechanism significantly differs from that of charged microgels, in which the free counterions surrounding the particles are controlling the deswelling behavior[29,30]. The size reduction is accompanied by a very limited particle interdigitation. The experimentally observed aggregation phenomena were reproduced in simulations of particles interacting with a short-range attraction in addition to the mild, long-range repulsion associated with electrostatics. The origin of the attractive interaction, which is atypical for polyelectrolyte brushes, could be attributed to base stacking interactions. These become significant in the osmotic regime where dsDNA fragments are stretched, and the blunt-ends of neighbor brushes face each other at a short distance. These blunt-end interactions could be tuned by acting on chain conformations through the addition of monovalent salt.

**RESULTS AND DISCUSSION**

**Experiments: Effect of packing on the brush size**

We present in this section the evolution of the conformation and interactions of increasingly packed SPBs extracted from the analysis of the radial distribution function $g(r)$ of quasi-2D dispersions of dsDNA-coated Polystyrene (PS) particles. A sketch illustrating the quasi-2D confinement is shown in Fig.1A: particle dispersions are contained in a channel formed by a microscope slide and a glass coverslip, separated by 10 $\mu m$ using spacers. The radial distribution function $g(r) = N(r)/(2\pi n r \Delta r)$, with $N(r)$ the number of particles in a thin shell of thickness $\Delta r$ at distance $r$ from a selected particle and $n = \langle N_p/A \rangle$ the average particle number surface-density, was determined from particle coordinates extracted from bright-field microscopy experiments. This contrast technique was chosen to avoid possible damages of the DNA chains due to prolonged exposure to intense laser irradiation in fluorescence-based techniques using labeled DNA[31]. In addition, fluorescent labeling of the dsDNA chain ends can alter their interactions[32]. In bright field contrast only the PS cores are visible. Dense dsDNA brushes were formed by grafting $f \approx 10^5$ dsDNA strands of length equal to 10 kilobase-pairs (kbp) on PS particles with radius $R_{PS} = 0.49 \mu m$. As shown in previous work[33,34], in water solution without any added salt the dsDNA chains assume a fully stretched configuration resulting in a brush thickness equal to the contour length $L_C = 3.4 \mu m$, corresponding to a brush-core size ratio $L_C/R_{PS} \approx 6.9$, i.e. a star-like architecture[35]. Fig. 1A shows an exemplary portion of an image of a dispersion, with indication of the overall size of the particles. Details of the synthesis, the preparation of dispersions and the quasi-2D confinement of the system are reported in the "Methods" section. The $g(r)$ functions of systems with increasing packing fraction $\eta = n\pi\sigma_0^2/4$, shown in Fig. 1B, evidence significant structural variations, indicated by changes in the number of the observed peaks, their height and position. Note that $\eta$ is calculated using the particle diameter in dilute solution $\sigma_0 = R_{PS} + L_C$ and thus reflects the increase in particle number density without accounting for the particle shrinking discussed later. A clear non-monotonic variation of the height of the first peak with increasing $\eta$ is evident and will be discussed later together with variations of the local order parameter. Here we focus instead on the variations of the position of the first peak of the $g(r)$, $r_p$, indicated by the dashed lines in Fig. 1B, which represents the shortest inter-particle distance. For monodisperse hard spheres, this quantity presents a minimum value equal to the particle diameter when particles are in contact.

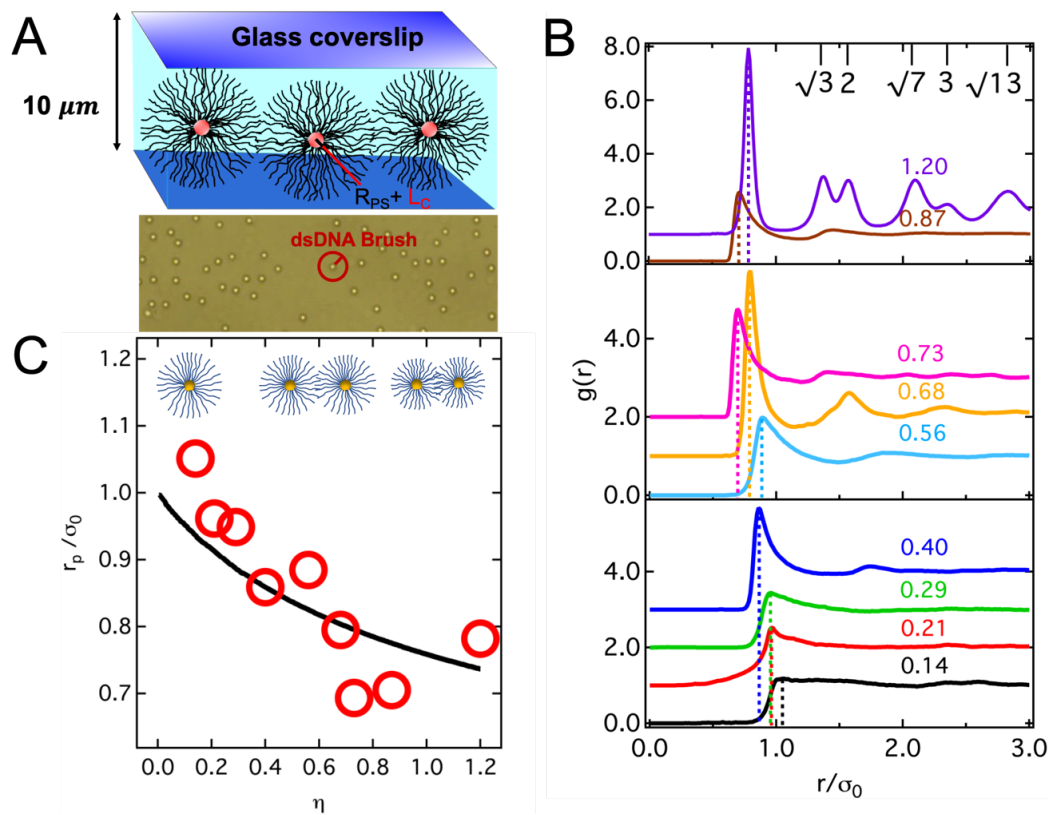

**Figure 1.** Experimental setup and Analysis of the Radial Distribution Function. **(A)** Top: Sketch of the experimental setup, showing dsDNA coated particles confined between two microscope coverslips separated by 10 μm spacers. The core ($R_{PS}$) and brush ($L_C$) sizes are indicated. The coverslips were coated with a hydrophobic material to avoid particles sticking to the glass. Bottom: Exemplary portion of a bright field microscopy image of a dispersion of dsDNA-coated colloids with packing fraction $\eta = 0.14$. The PS cores are visible. The overall size of the particles is indicated by the red circle. **(B)** Radial distribution functions g(r) of dispersions with different packing fractions $\eta$, as indicated. In each panel curves with larger packing fractions have been shifted vertically by 1 with respect to the previous curve for clarity. Dashed lines indicate the position of the first peak for each curve. In the top panel the expected positions of the peaks and the corresponding values of the ratios $r_i/r_1$ for a 2D hexagonal lattice are reported, with $r_i$ and $r_1$

the positions of peak $i$ and 1, respectively. **(C)** Position of the first peak of g(r) in units of the particle diameter in dilute conditions, as a function of packing fraction. Symbols: experiments, solid line: theory. The progressive size reduction as a function of packing fraction is represented in the cartoon.

For dense dispersions of soft particles, $r_p$ can be smaller than the particle diameter measured in dilute solution due to particle interpenetration, compression (shrinking) or deformation (and combinations of them). It can be therefore used to measure morphological changes of the particles with increasing crowding. Starting from $r_p \approx 1.05\sigma_0$ for $\eta = 0.14$, increasing packing fraction the inter-particle distance decreases monotonically until $\eta = 0.40$, for which $r_p \approx 0.88\sigma_0$. Interestingly, for $\eta = 0.56$ the position moves back to a slightly larger value, indicating a re-entrant behavior. For $\eta = 0.68$, the value of $r_p$ decreases again. Note that for this sample also the position of the second peak shifts to significantly smaller values compared to the previous sample, indicating a sudden compaction of the particle neighborhood. The minimum value of $r_p$ is registered for $\eta = 0.73$, while for $\eta = 0.87$ it increases slightly and significantly more for $\eta = 1.20$. For this value of $\eta$, the system crystallizes into a 2D hexagonal lattice, as demonstrated by the position of the peaks of the $g(r)$ (Fig. 1B). The decreasing, non-monotonic trend of $r_p$ as a function of $\eta$ is reported separately in Fig. 1C; Overall, the values of $r_p$ lie between $1.05\sigma_0$ and $0.7\sigma_0$. The fact that $r_p$ values are, for the majority of samples, smaller than the particle diameter and decreasing with increasing $\eta$ indicates that the dsDNA brushes were shrinking, interpenetrating and/or mechanically deforming. Interpenetration could be excluded in this case according to previous work[33]. In the next section, we present a theoretical model that explains the physical origin of the observed decrease of $r_p$ in terms of particle shrinking which results from the osmotic pressure generated by absorbed, non-condensed counterions. We will also show in the last section that the particle dynamics even at large packing fractions never fully arrest. This finding further supports shrinking as the origin of the reduction of $r_p$: if an increasingly large number of particles would be able to pack through deformation, particle movements should be strongly suppressed. Shrinking, deformation and interpenetration due to crowding have been intensively studied in microgel suspensions[30,36–42]. For neutral microgels three regimes were recognized: A first regime, below space filling, where no significant shrinking, interpenetration or deformation are observed; a second, above space filling, in which particle deformation and interpenetration occur; a third at even larger packing fractions in which deformation and interpenetration saturate and particles shrink (de-swell)[39,41]. Instead, for ionic microgels shrinking was found to be the main mechanism acting above space filling[42]. Note that the behavior of the system investigated here is clearly different from the two cases just discussed: a significant shrinking is observed well before space filling and interpenetration is negligible also at large packing fractions.

**Theory: Modeling the dsDNA brush configurations as a function of packing**

The primary dependence of the interaction diameter σ of SPBs on their packing fraction $\eta$ is a monotonic decrease, represented, roughly, by the solid line in Fig. 1C. Since the SPBs are complex macromolecular aggregates, their conformations and interactions depend crucially on a diverse variety of physical parameters, bringing forward their hybrid polymer/colloid character. An understanding of the effective interactions between SPBs requires analysis of the conformations of the same, which result from a minimization of a suitable free energy $F$, as we elaborate below. We build a cell model of a SPB of which the geometry is schematically illustrated in Fig. 2A. The brush consists of a hard-core having a radius $R_{PS}$, surrounded by a brush of thickness $L$, which consists of $f$ PE chains comprising $N$ monomers each. Correspondingly, for each chain there are $N$ monovalent counterions, which are contained in a Wigner-Seitz cell with radius $R_W$, related to the overall packing fraction $\eta$ by the requirement that a single SPB is contained within the volume of one cell. Since the experiment has shown that the size of SPBs can be highly density dependent, we explicitly differentiate between the density dependent brush size $R = R_{PS} + L$, and the brush size in unperturbed dilute conditions $R_0 = \sigma_0/2 = R_{PS} + L_C$, where $L_C$ stands for the brush height at infinite dilution. We assume that the SPB is dissolved in a solvent with electric permittivity $\varepsilon$ at temperature $T$. Because of the high bare charge of the DNA fragments, many of the counterions will be condensed along them. The number of which, $N_1$, can be approximated with the Manning parameter $\xi$, which we define as the ratio of the Bjerrum length, $\lambda_B = \frac{e^2}{4\pi\varepsilon k_B T}$, and the distance $b$ between each charge, giving $\xi = \lambda_B/b$, $k_B$ being Boltzmann's constant. Using this parameter, we estimate that the number of condensed counterions is $N_1 = \left|\frac{Q_{\text{bare}}}{e}\right|\left(1 - \frac{1}{\xi}\right)$, where $Q_{\text{bare}} = -efN$ is the total bare charge of a brush [43]. With this estimation we presume that the counterions condense in such a way that there remains only one net charge per Bjerrum length. Here, we neglect the effect other nearby chains exert because their electrostatic interactions are screened by the counterions in the brush. The remaining $fN - N_1$ counterions are subdivided into two populations $N_2$ and $N_3$, representing those that are free to move within the brush, and those that are free outside of the brush. In order to find the number of free counterions within the brush ($N_2$) and outside of the brush ($N_3$), as well as the brush thickness $L$, we set-up a variational free energy $F(N_3, L)$, and minimize it to find the equilibrium values of $N_3$, and $L$. The remaining population is now easily calculated with the charge-neutrality condition $N_1 + N_2 + N_3 = Nf$. We explicitly minimize for the brush thickness $L$ as well to be able to qualitatively capture the significant decrease of the location of the first peak of the radial distribution function $g(r)$ that was found in the experimental system shown in Fig. 1C.

In this free energy $F$, we include six contributions and express it as:

$$F = U_\text{H} + F_\text{el} + F_\text{Fl} + S_2 + S_3 + F_\text{p}. \tag{1}$$

We present more information and explicit calculations in the Supporting Information, limiting ourselves to a more concise description in what follows.

The first contribution ($U_\text{H}$) approximates the electrostatic energy modelling the Coulombic interactions between all the charged PE-monomers and counterions. We use a Hartree type expression

$$U_\text{H} = \frac{1}{8\pi\varepsilon} \iint drdr' \frac{\rho(r)\rho(r')}{|r-r'|} \tag{2}$$

where $\rho(r)$ is the expectation value of the total charge density resulting from the sum of the counterion charges and the charges on the PE chains, and where the integrations run over the entire Wigner-Seitz cell. Theory, simulation and experiment agree that the chains of isolated, dense PE brushes in a salt-free environment are completely stretched, meaning that the charged monomer density falls of as $r^{-2}$ inside the brush[33,44,45]. Because counterions are inclined to neutralize the charged monomers, we assume that the distribution of counterions within the brush also has this functional form. Furthermore, we model the free counterions outside the brush as homogeneously dispersed. The charge density resulting from the sum of the counterion and the monomer density is now given by

$$\rho(r) = \begin{cases} 0 & r < R_{PS} \\ \frac{-eN_3}{4\pi L r^2} & R_{PS} \leq r \leq R \\ \frac{3eN_3}{4\pi(R_W^3 - R^3)} & R < r < R_W \end{cases} \qquad (3)$$

in which the pre-factors ensure charge neutrality.

The second term ($F_{el}$) in the free energy models the entropic elasticity of the PE-chains, which adds a penalty for a highly stretched chain configuration[45]

$$\frac{F_{el}}{k_B T} = \frac{3fL^2}{2Nb^2} \qquad (4)$$

Here $b$ is the equilibrium length of the bonds between chain monomers.

The third ($F_{Fl}$) is a Flory self-avoidance term that models the excluded-volume interactions between the chain monomers. We choose to set the excluded volume equal to the volume of a monomer with radius $R_m$ and obtain[44].

$$\frac{F_{Fl}}{k_B T} = \frac{f^2 N^2 R_m^3}{2L^3} \qquad (5)$$

The form of this contribution to the free energy is strictly only applicable if the radius of the central colloid is much smaller than the total brush radius $R_{PS} \ll R$. Similarly, we implicitly assume that the chain monomer density within the brush is homogeneous, meaning that the PE chains can fully explore the volume of the brush and are not attached to the central colloid. However, because we are mainly interested in predicting the way in which the brush size $L$ changes with a change in the relevant parameters, we expect that these simplifications do not disqualify our findings.

The next two terms model the entropic free energy $S_2$ of the non-condensed counterions within the brush and that of the free counterions outside the brush, $S_3$. We leave out the entropy of the condensed counterions as it will drop out in the minimization since the number of these counterions is kept constant. However, we do take into account that the presence of the PE chains limits the available free volume to the counterions in the brush. Defining the local number densities $n_i(r)$, $i \in \{2,3\}$ and the counterion diameter $d$, we can estimate the entropic contributions to the free energy with

$$S_i = k_B T \int_{V_i} dr\, n_i(r) ln(n_i(r)d^3) \qquad (6)$$

in which we omit the usual characteristic length-scale term, as it will yield a constant to the free energy[45].

The final contribution ($F_p$) to the free energy takes into account the effects of the surrounding SPBs in a concentrated solution on the size of any given SPB. Following the arguments put forward for the related case of star polymers[46], we introduce $F_p$ as the free energy cost of insertion of a SPB of radius $R$ in a concentrated solution of the same. Such an insertion results into the expulsion of the remaining SPBs from a region of size $R$ and the associated free energy cost can be estimated as the product between the volume taken up by the SPB, $R^3$, and the osmotic pressure $\Pi(R_W)$ of the remaining solution at packing fraction $\eta$, parameterized through the Wigner-Seitz radius $R_W$. This osmotic pressure, in turn, is dominated by the trapped counterions[46] and is estimated by the product of the number of entropically active counterions $N_2$ in each brush, the thermal energy $k_B T$, and the number density of the brushes $\propto R_W^{-3}$. Summarizing, we obtain

$$\frac{F_p}{k_B T} = \frac{N_2 R^3}{R_W^3} \tag{7}$$

and can only be considered in the limit that $N_2 \gg N_3$, which we justify later in this section.

The role of the various terms is antagonistic: some of them favor large SPB sizes whereas others would favor the shrinking of the same, thus their competition leads to a value that minimizes the total free energy. The sum of these contributions is numerically minimized with respect to $L$ and $N_3$. In Figs. 2C-D, we show for miniature brushes ($N = 40$, $R_{PS} = 10$nm) the effect of varying the functionality $f$ and the SPB density on the osmotic power of the brushes, *i.e.*, on the fraction of counterions the brush absorbs. The name miniature is used to indicate that the simulated brushes are a scaled version of the experimental system presenting the same grafting density (but smaller fragment length, see Fig. 2B). We validate the results of this procedure with coarse-grained molecular dynamics simulations of these small brushes (see the Supporting Information), finding excellent quantitative agreement between the theoretical cell model and simulations for the fraction of the non-absorbed counterions (Figs. 2C-D). We show that as the functionality increases, a decreasing fraction of counterions manages to escape the brush (Fig. 2C). This is due to the increased relative influence of the electrostatic energy.

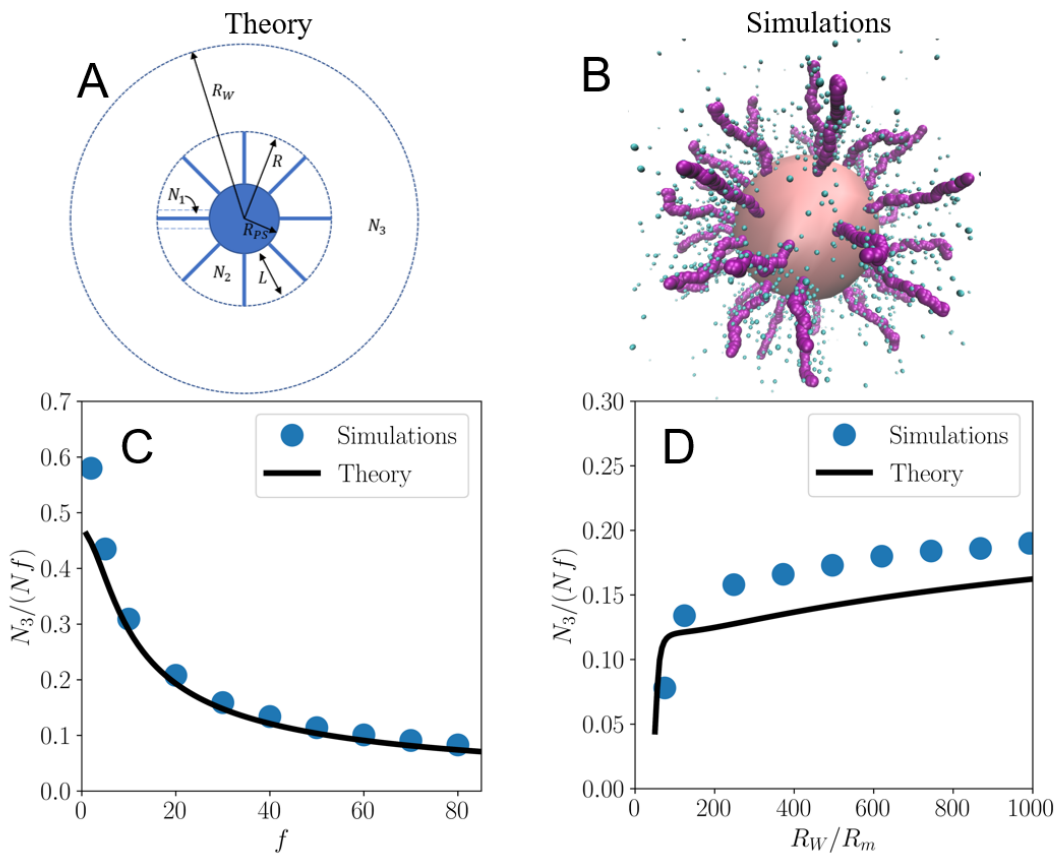

**Figure 2.** Theoretical and simulation models, fraction of non-condensed counterions **(A)** Schematic model of a SPB of total radius $R = R_{PS} + L$, enclosed in a spherical Wigner-Seitz cell of radius $R_W$. $N_1$, $N_2$ and $N_3$ are the numbers of condensed, non-condensed and free counterions, respectively. **(B)** Snapshot of coarse-grained MD simulation of miniature SPB ($N = 40$, $R_{PS} = 10$nm), which presents the same grafting density as the experimental system. **(C, D)** Fraction of counterions $N_3/Nf$ that escape the SPB, where $f$ is the functionality and $N$ the chain length. Figure (C) shows this fraction as a function of functionality $f$ and (D) as a function of Wigner-Seitz radius $R_W$ scaled with the radius $R_m$ of a chain monomer. For more information on the simulations of the miniature brushes, we refer to the Supporting Information.

Since the functionality of the experimental brushes is rather large ($\sim 10^5$), we expect these to be highly osmotic, releasing very few counterions. Indeed, using the geometric parameters of the large experimental brushes, our cell model predicts that the fraction of released counterions is of the order of $10^{-5}$, indicating that virtually all available counterions are being absorbed. On the other hand, we see that an increase in the cell size $R_W$, related to the volume fraction by $\eta = R_0^3/R_W^3$, tends to decrease the osmotic power of the brushes. This is to be expected, of course, since an increase in the available volume for each brush increases the entropy of released counterions.

Applying the model to brushes with the same geometry as that from the experiments results in a concentration-dependent size given by the solid line in Fig. 1C, which describes very well the progressive reduction of the position of the first peak of the experimental $g(r)$. For the large experimental brushes, we find that especially the contribution $F_\text{p}$ has a very significant influence on the density dependence of the brush size. Even though the model captures well the experimentally observed density-driven shrinkage of the brushes, it does not manage to accurately predict the absolute size of the experimental brushes, underestimating it by roughly 40%. On the other hand, such deviations are not unusual in scaling-type theories that seek to establish general trends and regimes and do not aim at detailed quantitative accuracy.

Even though the SPBs shrink as the local colloid density increases, at constant density the effective interactions between two such particles can be expected to be very strongly repulsive as they start overlapping. In particular, Jusufi *et al.* showed that the effective interactions between colloidal particles similar to SPBs scale linearly with the number of absorbed counterions[45]. To confirm that these findings extend to our system as well, we perform a similar analysis for miniature SPBs of which we present the results in the Supporting Information. Since the number of absorbed counterions in the experimental brushes must be of the order of $Nf \approx 10^9$, we conclude that even small brush overlaps are penalized with energies orders of magnitude higher than those available for thermal fluctuations. In short, our model prohibits highly osmotic SPBs from (significantly) interdigitating, in agreement with experimental findings in previous work[33].

**Experiments: Aggregation and Re-entrant ordering**

While the theory developed in the previous section identifies the physical origin of the progressive size reduction of the brushes, the experimental data show an additional, non-monotonic trend of the position of the first peak, with significant deviations from the theoretical predictions in the interval $0.7 < \eta < 0.8$. This discrepancy suggests that the monotonic size reduction predicted by theory as an effect of the osmotic pressure of the absorbed, non-condensed counterions, is not sufficient to capture the entire experimental phenomenology, at least when only repulsive interactions between the brushes are considered. To better understand the physical origin of the non-monotonic experimental trend, we report in Fig. 3 the packing fraction dependence of two additional structural parameters, the height of the first peak of the $g(r)$, $g_p$, and the average six-fold order parameter, $\langle \Psi_6 \rangle = \langle \frac{1}{N} \sum_{i=1}^{N} \Psi_6^i \rangle$, with $N$ the total number of particles in one image of the sample, where $\Psi_6^i = \frac{1}{N_b} \sum_{j=1}^{N_b} e^{i6\vartheta_{ij}}$ is the six-fold order parameter of particle $i$ with $\vartheta_{ij}$ the relative orientation angle between particles $i$ and $j$ and $N_b$ the number of neighbors of particle $i$. The brackets $\langle \ \rangle$ indicate an average over all images of the sample.

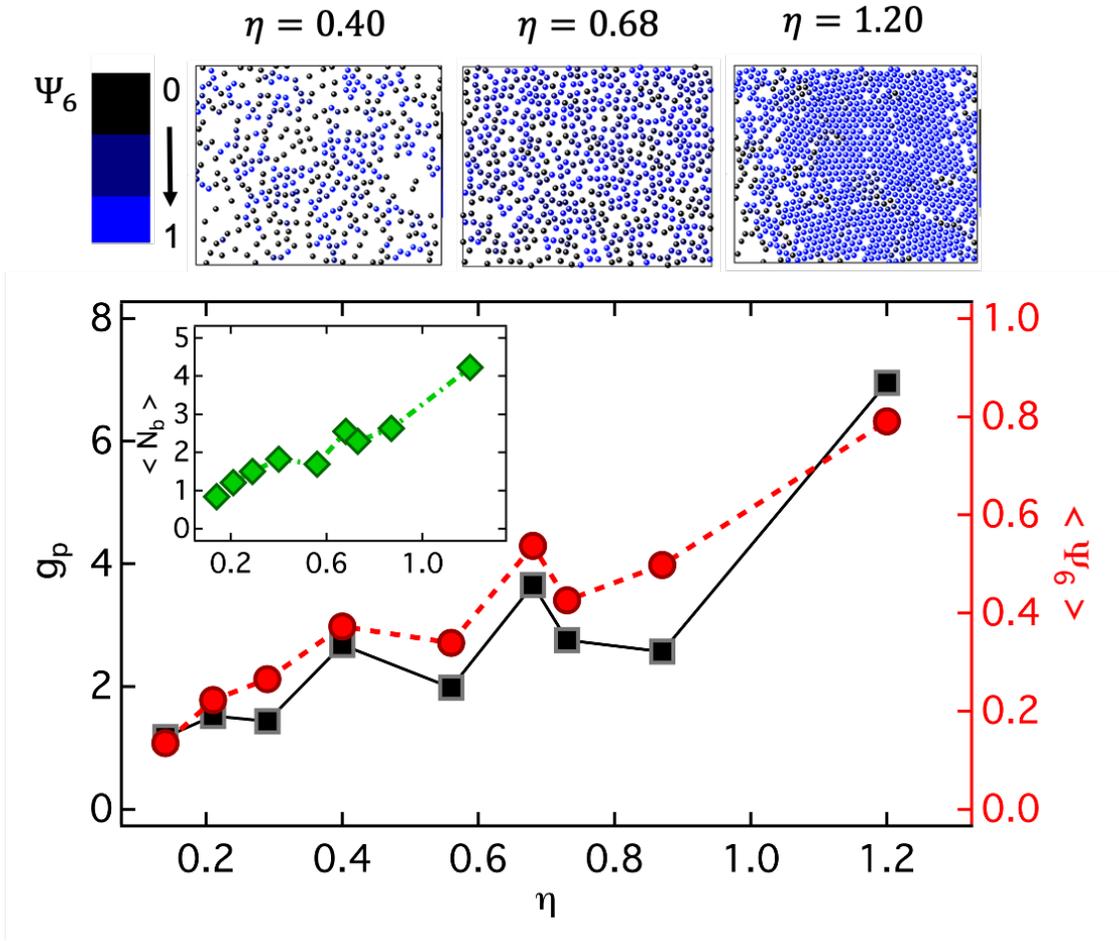

**Fig. 3.** Structural parameters to investigate the degree of ordering. **(Bottom)** Height of the first peak of the g(r), $g_p$ (squares) and average six-fold order parameter $\langle \Psi_6 \rangle$ (circles), as a function of packing fraction $\eta$. Inset: Average number of neighbors per particle, $\langle N_b \rangle$, as a function of packing fraction $\eta$, same x-axis as the main plot. **(Top)** Images corresponding to exemplary renderings were obtained from coordinates for the samples indicated by the arrows. Particles are colored according to the value of $\Psi_6$.

Two particles were considered neighbors when the distance between their particle centers was smaller than the diameter plus half the distance between the first maximum and the first minimum of the $g(r)$. Both $g_p$ and $\langle \Psi_6 \rangle$ show a similar non-monotonic trend which, on top of a progressive increase as a function of $\eta$, shows the presence of two local peaks. One is observed for $\eta \approx 0.40$ and the second for $\eta \approx 0.68$. These peaks indicate for the corresponding samples the presence of structures with a larger degree of local order and are visualized in the representative renderings of the samples shown in Fig. 3, which were obtained using particle coordinates from particle-tracking (representative renderings of all samples can be found in the Supporting Information, Fig. S1). For the sample with $\eta \approx 0.40$ one can see that chain-like structures are present with a simultaneous emergence of density inhomogeneities (crowded regions and voids), indicating the presence of attractions in the effective brush-brush interaction potential. These attractions, however, are neither broad nor deep enough to bring about macroscopic phase separation (liquid-gas) in the SPB-solution, leading rather to the formation of finite-size clusters only. The linear geometry of the structures is also confirmed by the average number of neighbors for each particle which is $\langle N_b \rangle \approx 2$ (Fig. 3, inset). The length of the chains presents a broad distribution and also isolated particles are present. For the sample with $\eta \approx 0.68$, the aforementioned features persist, albeit with suppressed density inhomogeneities with respect to $\eta \approx 0.40$, and the average number of neighbors increases to $\langle N_b \rangle \approx 3$ (Fig. 3, inset). The degree of local order within the aggregates is pronounced, as confirmed by the value of the average six-fold order parameter, $\langle \Psi_6 \rangle \approx 0.54$. Interestingly, for $\eta > 0.68$ $g_p$ and $\langle \Psi_6 \rangle$ (and also $\langle N_b \rangle$ to a minor extent) decrease first and then increase again, indicating a re-entrant order-disorder transition. The observed trend can be associated with that of $r_p$, which for $\eta = 0.73$ shows the minimum value, indicating strongest size reduction, and increases for larger values of $\eta$. The sudden disordering can be thus associated with the size reduction and the successive increase of order with a re-swelling.

We already commented on the fact that the size reduction registered for these samples deviates from the monotonic trend predicted by assuming purely repulsive interactions and we additionally noted that pronounced aggregation is observed for these packing fractions. This leads us to conclude that the pronounced particle shrinking responsible for the reentrant transition might be associated with the strong local packing induced by aggregation. These aggregation phenomena can be explained if an attractive interaction between the brushes is present. We will demonstrate that introducing a model short-range attraction between particles/brushes the experimental re-entrant ordering can be qualitatively reproduced; the origin of this attraction can be attributed to blunt DNA ends, and its effect can be suppressed by salinity. Note, finally, that the formation of a hexagonal lattice for the highest packing fraction $\eta = 1.20$ is confirmed by the large value of $\langle \Psi_6 \rangle \approx 0.8$.

**Experiments: Testing the origin of the short-range attraction by acting on DNA conformation**

The experimental observation of finite-size clusters and density inhomogeneities discussed in the previous section indicates the presence of a short-range attraction of moderate intensity between the dsDNA brushes. We speculate that DNA blunt-end base stacking is the origin of this effective attractive interaction. Our interpretation is based on the following arguments: Previous experiments[33] and findings in this work indicate the absence of significant interdigitation at large packing fractions and a stretched configuration of the dsDNA fragments within the brushes in the absence of added salt. We can therefore foresee that when two neighbor brushes are in contact and the DNA fragments are stretched, a large number of blunt ends on the two sides will face each other and will be separated by a short distance. Recent experimental and theoretical work on the assembly of DNA nanostructures[21–23,47,48] showed that when a large number of complementary

DNA blunt ends lie at sufficiently small distance from each other, stacking assembly is observed. We propose that this mechanism could be at the origin of the attractive interactions between compressed and densely packed DNA brushes. It was found that the attractive interaction per base contact amounts to a few $k_BT$[22]. Upon the SPBs approaching contact, blunt-end-pairs are exposed to such an attraction, while at the same time experiencing a weak electrostatic repulsion from the other brush, which is of the order of a few $k_BT$ itself[46]. The resulting effective interaction between two SPBs could be thus estimated to be of the order of the thermal energy. A fundamental assumption of our hypothesis is the stretched configuration of dsDNA fragments due to the presence of a large fraction of counterions absorbed within the brush in the absence of salt. To test our hypothesis, we performed a similar analysis of the structural evolution of the dsDNA brushes adding 25mM of NaCl to the dispersions. We report in Fig. 4A exemplary images of samples with comparable particle number density for the system in deionized water and with addition of 25mM NaCl. For the sample in deionized water, which corresponds to $\eta \approx 0.68$, in selected regions like the one reported in Fig. 4A (left) one observes a pronounced heterogeneous structure with aggregates and local ordering within the aggregates. For the sample with 25mM NaCl we did not find instead heterogeneous regions and the structure is generally homogeneous (Fig. 4A, right). This suggests an important change of the effective interactions in the sample with 25mM salt. This pronounced difference is confirmed by the comparison of the trend of $g_p$ as a function of $\eta$ for the two systems (Fig. 4B): No peaks are visible for the system with added salt (The corresponding $g(r)$ are reported in the Supporting Information, Fig. S2).

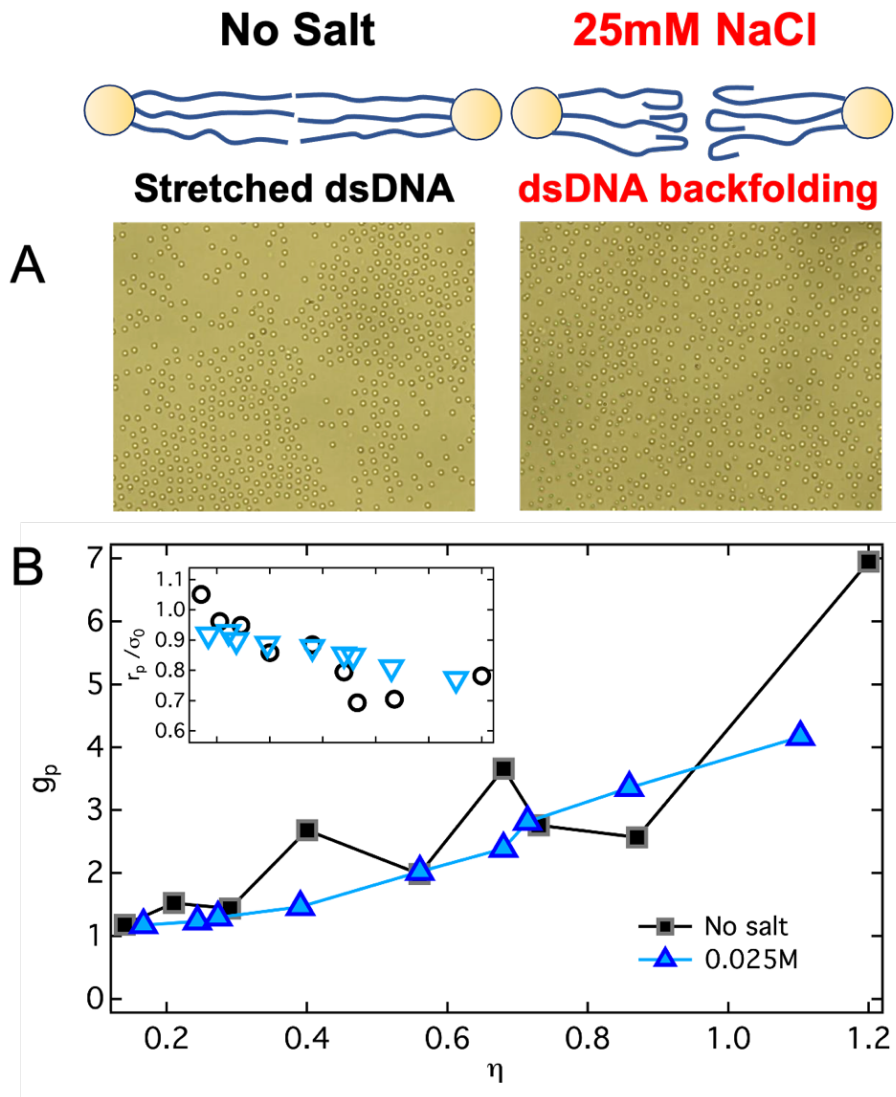

**Fig. 4.** Changes in the dsDNA configuration lead to the disappearance of reentrant ordering. **(A)** Images of samples with $\eta \approx 0.68$ for systems without any added salt (left) and with 25mM NaCl (right). **(B)** Height of the first peak of the $g(r)$, $g_p$, as a function of packing fraction $\eta$, for the system without any added salt (same data as in Fig.3) and for the system with 25mM salt content, as indicated. Inset: Corresponding peak position $r_p$ as a function of $\eta$, same range as in the main plot. At the top of the figure, we show a schematic representation of the changes in the configuration of a few exemplary dsDNA fragments for the system with and without added salt.

Additionally, also the position of the first peak decreases in this case smoothly, different from the case without added salt (Fig. 4B, inset). We speculate that while addition of monovalent salt should increase the strength of blunt-end interactions[22,47], at the same time it strongly affects the spatial configuration of the dsDNA fragments: Previous work in [33] showed that a significant backfolding (sketch in Fig. 4) of the ends of the fragments occurs. This implies that the probability that blunt ends from neighbor brushes face each other drastically reduces, and thus the effective attraction between brushes decreases.

**Simulations: Aggregation and re-entrant ordering in a system with competing short-range attraction and mid-range repulsion**

Based on the experimental evidence on the presence of additional, short-range attractions originated by blunt-end interactions, we postulated an effective repulsive potential that includes additional such attractions and confirm that it brings about the experimentally observed features. In particular, we performed Monte Carlo simulations of colloidal particles in two dimensions, interacting with the following generic pair interaction:

$$V(r) = 4V_1 \left[ \left(\frac{\sigma}{r}\right)^{200} - \left(\frac{\sigma}{r}\right)^{100} \right] + V_2 \frac{e^{-r/\lambda}}{r/\lambda} \qquad (8)$$

The first term is the Lennard-Jones 100-200 potential, modeling a strong repulsion of hard-core-like spheres of diameter $\sigma$, followed by a short-range attraction.

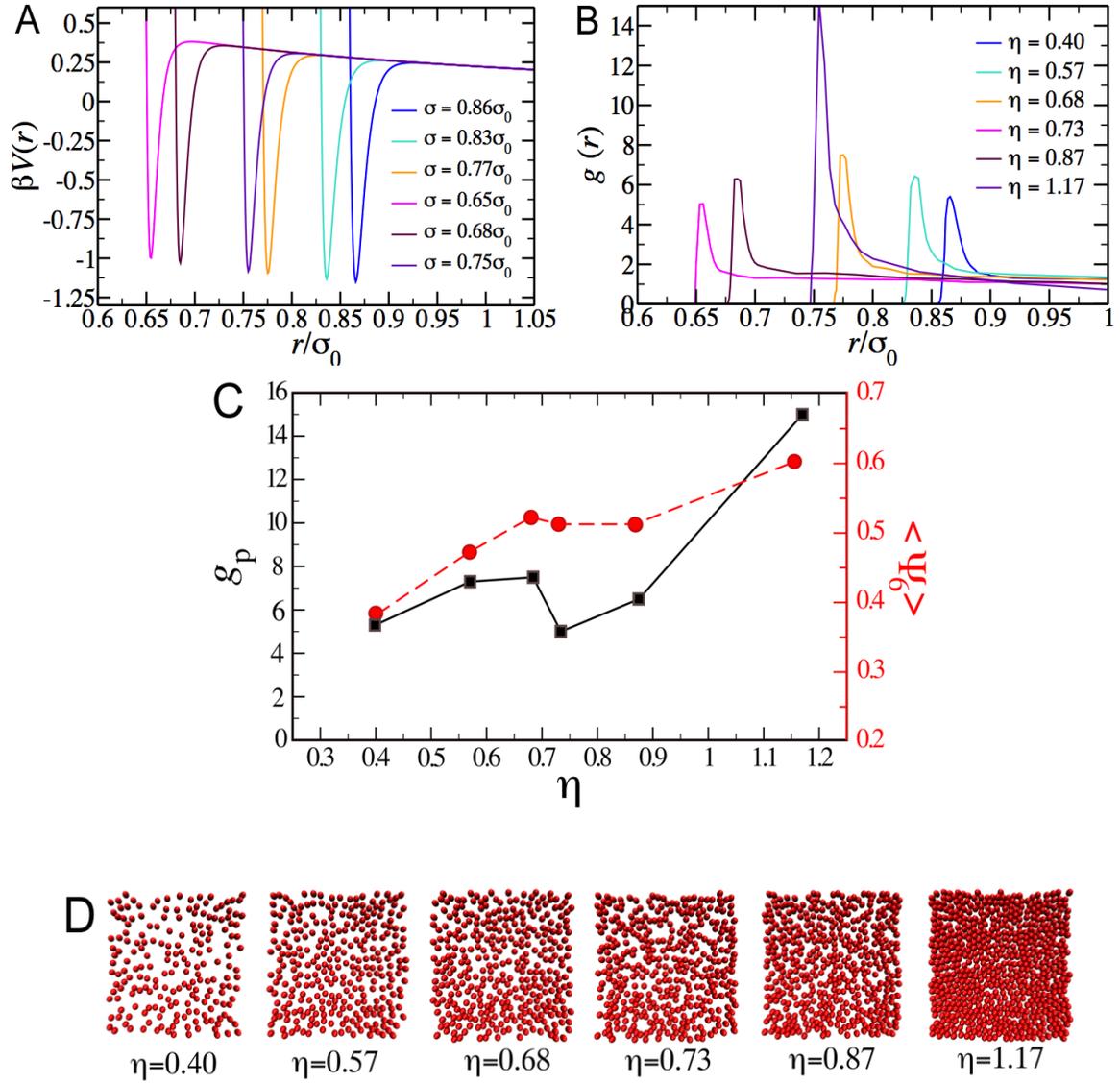

**Fig. 5.** MC simulations of a system with competing attractive and repulsive interactions confirm reentrant ordering. **(A)** Interaction potential with parameters: $V_1/k_BT = 1.43$; $V_2/k_BT = 0.28$; $\lambda/\sigma_0 = 1.5$, and variable $\sigma$ as indicated in the legend; **(B)** simulated g(r) for various packing fractions, where the color code matches the scheme in (A); **(C)** the trend of the height of the first peak (black symbols) and the order parameter (red symbols) as a functions of packing fraction; and **(D)** corresponding simulation snapshots.

The justification for such strong repulsion which prohibits particles from interdigitation, is found in the work of Jusufi *et al.* who showed that the effective interactions between colloidal particles similar to SPBs scale linearly with the number of absorbed counterions[45]. Since the number of the absorbed counterions in the experimental brushes must be of the order of $Nf \approx 10^9$, we conclude that even small brush overlaps are penalized with energies that are orders of magnitude larger than thermal fluctuations. This assumption is corroborated by the experimental findings of[33] where it was shown that dsDNA-coated colloids densely packed on a 2D lattice are resilient to mutual interpenetration of their charged coronas. Moreover, the Lennard-Jones term in Eq. 7 features an attractive well with a range corresponding to a fraction of $\sigma$. This short-ranged attraction is used to model interactions between dsDNA fragments when particles approach each other to close proximity, caused by the blunt DNA-end, as discussed in the previous section. The second term in Eq. 7 is of a repulsive Yukawa form that models a weak residual electrostatic repulsion between the almost fully neutralized SPBs.

The potential of Eq. 7 has been used to investigate aggregation phenomena in 3D colloidal systems, showing that the competition between short-range attraction and mid-range repulsion drives the formation of aggregates. The choice of the potential parameters, $V_1$, $V_2$ and $\lambda$, determines the morphology of the aggregates[49,50]. In the simulations based on the potential of Eq. 7 the experimentally observed size reduction of the particles was included by using the values of the experimental particle diameter as a function of the packing fraction $\eta$ in Fig. 1C. Potentials with different sets of $V_1$, $V_2$ and $\lambda$-values were generated. The potential generating the $g(r)$ that shows reasonable, semi-quantitative agreement with experiments was chosen as the most representative of the interactions in the experimental system.

The pair potential is reported in Fig. 5A and was obtained for $V_1/(k_B T) = 1.43$; $V_2/(k_B T) = 0.28$; $\lambda/\sigma_0 = 1.5$. Although we have not attempted a microscopic derivation for the values of the parameters used, it is possible to offer plausibility arguments for the resulting values on the basis of physical argumentation. The parameter $V_1/(k_B T)$ sets the scale of the attraction, which, as suggested in the previous section, is caused by end-to-end stacking of the dsDNA blunt ends and is expected to be of the order of the thermal energy. The obtained value of $1.43 k_B T$ is in very good agreement with this expectation and supports our interpretation about the origin of the attractive interaction. On the other hand, the value of the parameter $V_2$ is set by the overall SPB-charge, which is very low for osmotic brushes and thus a small value results. Finally, the screening length $\lambda$ is set by the radius of the Wigner-Seitz cell, which is somewhat larger than the brush size in the concentrations under consideration.

The obtained particle configurations were used to determine the corresponding $g(r)$ which are presented in Fig. 5B for the investigated values of $\eta$. A non-monotonic behavior of the height of the first peak, $g_p$, for packing fractions in the range $0.7 < \eta < 0.8$, Fig. 5B, is observed, in qualitative agreement with the experimental findings. The re-entrance is also reflected in the non-monotonicity of the six-fold order parameter $\langle \Psi_6 \rangle$, Fig. 5C. Finally, the snapshots of the simulated systems (Fig. 5D) show the presence of chain-like structures and aggregates comparable to those found in experiments. The simulations thus confirm that attractions, which induce aggregation, are a key ingredient to explain the re-entrant ordering phenomenon. Therefore, we conclude that re-entrant ordering is determined by two mechanisms: The formation of aggregates and size reduction due to deswelling. At larger packing fractions, particles get more ordered and progressively shrink; at the same time, attraction induces formation of aggregates. When the aggregates become locally denser than the average packing fraction, a pronounced shrinking occurs, which leads to a sudden disordering. Aggregates are disrupted and the local packing decreases, allowing the particles to rearrange configurations and partially re-swell. Further increasing the packing fraction, the order increases again until crystallization occurs. It is interesting to note that a non-monotonic variation of $g_p$ at packing fractions in the range $0.7 < \eta < 0.9$ can also be found in Monte Carlo simulations in which a monotonic decrease of particle size similar to that predicted by theory is assumed, even though the agreement with experiments is poorer (results not shown). This suggests that there is a critical packing fraction, which depends on the degree of deswelling, at which a restructuring into a more disordered structure is needed to be able to pack additional particles. We remark also that the re-entrant behavior is only found in a very narrow range of potential parameters.

**Experiments: Dynamics as a function of packing fraction**

The structural variations observed with increasing packing fraction, and in particular the aggregation phenomena assigned to the interactions between dsDNA fragments of contacting brushes, should also affect the dynamical behavior of suspensions. In particular, the presence of particle aggregates should induce a slowdown of the average single-particle dynamics. To test our expectation, we determined the mean squared displacement (MSD) of samples with different packing fractions, $\langle \Delta r^2(\Delta t) \rangle = \left\langle \left( r_i(\Delta t + t_0) - r_i(t_0) \right)^2 \right\rangle_{t_0,i}$ where $r_i$ is the position of particle $i$, $\Delta t$ is the delay time, $t_0$ is the time during the particle trajectory, and $\langle \rangle_{t_0,i}$ indicates an average over all times $t_0$ and all particles $i$. We show the resulting MSD for several packing fractions in Fig.6. All trajectories were corrected for the possible presence of drift due to stage instabilities: despite this correction, the apparent super-diffusive behavior at very short times might be the result of residual drift contributions.

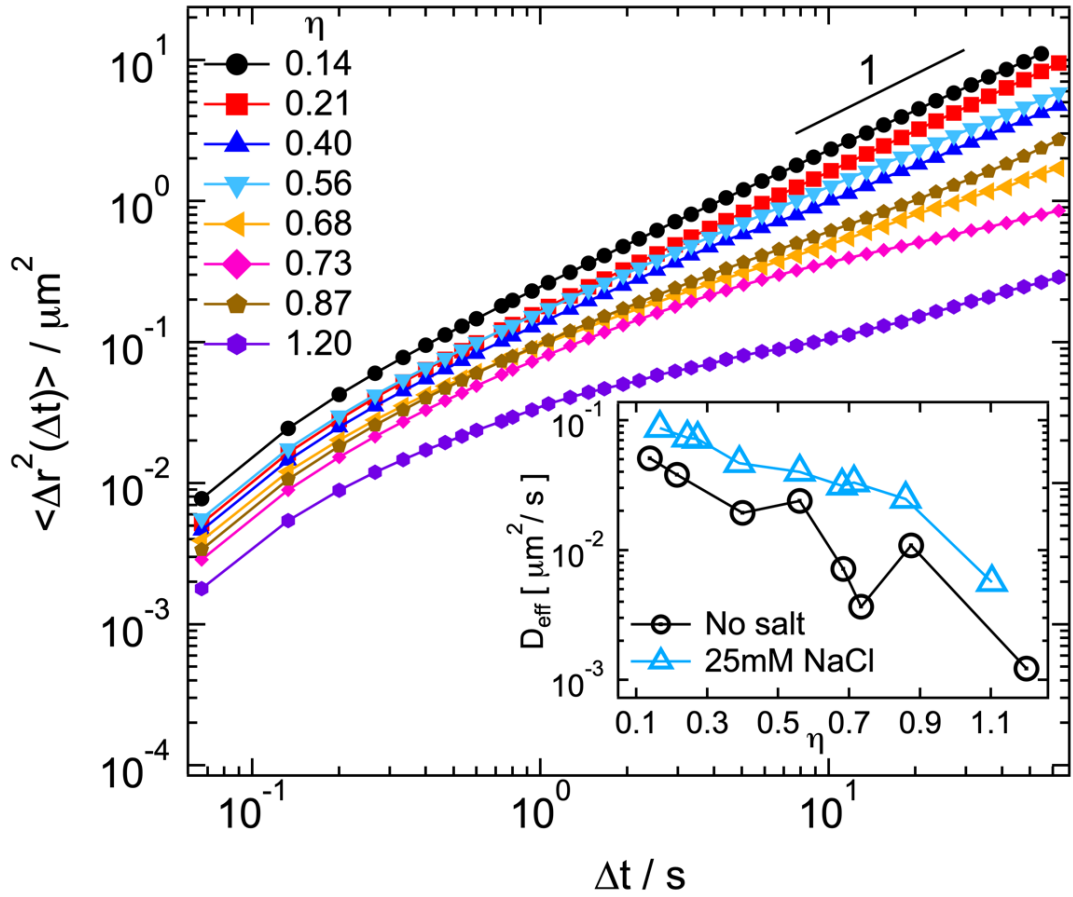

**Fig. 6.** Experimental mean squared displacements for samples with different packing fractions. Average mean squared displacement $\langle \Delta r^2(\Delta t)\rangle$ of dispersions of dsDNA-coated colloids with different packing fraction $\eta$, as indicated. Inset: Effective diffusion coefficient $D_{\text{eff}}$ extracted from the mean squared displacements, as a function of $\eta$ for the samples in the main panel and for the system with 25mM NaCl, as indicated.

Samples with $\eta \leq 0.56$ show approximately diffusive dynamics at long times, as indicated by the almost linear dependence of the MSD on $t$. Aggregation in the form of chain-like structures observed for $\eta = 0.40$ leads to a significant slowdown of the dynamics and smaller values of the MSD, while a slightly larger MSD is obtained for $\eta = 0.56$. This corresponds to the transition to a more uniform spatial distribution of particles with less aggregates. Similarly, a considerably smaller MSD is observed at $\eta = 0.68$ and $\eta = 0.73$: in particular the MSDs become sub-diffusive, in agreement with the formation of a large number of aggregates in which particle movements are suppressed. The MSD presents a larger value and a time dependence approaching that expected for diffusion for $\eta = 0.87$, while a significantly smaller and sub-diffusive MSD is obtained for $\eta = 1.20$. For the latter, the presence of a large number of crystalline regions is at the origin of the slow dynamics. The inset of Fig. 6 reports an effective diffusion coefficient calculated as $D_{\text{eff}} = \Delta r^2 \, (\Delta t \approx 55s)/4 \, \Delta t$, which summarizes the behavior illustrated above for the MSD, and which confirms the correspondence between the structural variations and the evolution of the dynamics. The MSDs of samples with 25mM NaCl are reported in Fig. S3 of the Supporting Information. Similar to what was discussed for the $g(r)$, also the dynamics of the system with salt show a smoother slowdown with increasing $\eta$, as shown in the inset of Fig. 6. The data of Fig.6 also confirm what was anticipated when discussing the possibility of brush shrinking and/or deformation with increasing packing fraction: For all samples, except the crystalline state for $\eta = 1.20$, the MSD shows a diffusive or moderately sub-diffusive behavior, indicating that dynamical arrest is not occurring even at large packing fractions. This supports the scenario of progressive shrinking of the brushes rather than deformation.

**Discussion**

We reported unconventional effects of packing on the morphology and interactions of thick, dense spherical dsDNA brushes in planar confinement. Combining experiments and theory we showed that the large number of free (entropically active) counterions absorbed within a dense brush in the osmotic regime produces a huge entropic pressure which leads to the progressive shrinking of neighbor brushes with increasing packing fraction. Interestingly, shrinking occurs without significant interdigitation of the brushes and starts well below space filling. Moreover, the absence of dynamical arrest, even for large packing fractions, suggests that shrinking prevents jamming and significant particle deformation. These findings mark a pronounced difference with the behavior of uncharged hairy colloids, where interdigitation is especially relevant[51], but also of neutral and charged microgel particles, in which deswelling occurs above space filling or even at larger packings[52]. As demonstrated in previous experimental and theoretical studies on planar polyelectrolyte brushes, a small degree of interdigitation plays a fundamental role in maintaining the lubrication between contacting brushes under high loads. SPBs find application as lubrication additives in biological environments[53,54]: Our study, indicating shrinking and small interdigitation of SPBs with increasing packing, suggests that low friction is expected between highly crowded brushes, a condition which is relevant for the applications mentioned above, in which SPBs dispersions are typically strongly confined. The lubrication between brushes is also supported by the dynamics of the system, which do not fully arrest even for highly crowded conditions.

The isotropic repulsive interactions derived in the theory for generic polyelectrolyte brushes do not entirely explain the structural evolution of the increasingly packed spherical dsDNA brushes. Aggregation phenomena in the form of chain-like structures and non-monotonic shrinking are observed experimentally and were reproduced in simulations by considering an additional short-range attractive interaction, in competition with electrostatic repulsion. We explained the origin of this attraction to base stacking forces between blunt ends of dsDNA fragments, which become particularly important when osmotic brushes are densely packed. In these conditions a large number of blunt ends from neighboring brushes lie at short distance and can attract each other, leading to an effective additional short-range attraction between the particles that drives assembly. Colloidal assembly exploiting DNA hybridization of single-stranded DNA or sticky ends has been largely investigated during the last years[26], mainly for assembling crystals[55,56] but also non-equilibrium gels[57]. However, it was found that assembling structures with a higher degree of complexity than those also obtained with more conventional colloids, and with a programmable approach similar to that used in DNA nanotechnology, is an extremely demanding task. As mentioned in the introduction, blunt-end base stacking has been shown to be especially powerful in DNA nanotechnology in combination with shape design[58,59], but is almost unexploited in colloidal assembly. We can foresee that engineering the DNA blunt-ends through careful design of the PCR primers, and control over their spatial distribution can allow the orthogonal programming of the directionality and strength of the interactions between dsDNA grafted colloids. The experimental realization of such novel patchy spherical DNA-based brushes will provide the basis for the development of new self-assembly platforms that combine directionality and sequence complementarity of DNA fragments. This may be used to guide the organization of colloidal materials with unique plasmonic[60,61] and photonic properties[62], thanks also to the possibility of easily changing the material and shape of the colloidal core[63,64], and thus the responsiveness to external fields.

**METHODS**

**Synthesis of DNA Star Polyelectrolyte Colloids and Dispersions' Preparation**

The procedure to obtain DNA star polyelectrolyte colloids was described in detail before [33]. Here, the main steps of the procedure are recalled. They can be summarized as follows: i) Synthesis of 10 kbp double-stranded DNA through the amplification of end-biotinylated fragments using Polymerase Chain Reaction (PCR). ii) Grafting of DNA chains to the streptavidin functionalized surface of polystyrene beads ($R_{PS} = 0.49$ μm). In step i) we amplified the end-biotinylated dsDNA fragments using a $\lambda$-DNA template (New England Biolabs) and a DNA polymerase enzyme contained in the Go Taq Long PCR Master Mix (Promega) and following the detailed PCR protocols accompanied with this product. End-functionalization of the dsDNA strands was achieved by the PCR using commercially synthesized and HPLC purified forward and reverse primers, modified at their 5'-ends (IDT). More specific, for the aforementioned linear dsDNA fragments the forward primer was 5′-biotinylated, including an extended 15-atom spacer TEG (Tetra-Ethylene-Glycol) in order to reduce steric hindrance and therefore increase the binding efficiency of the long dsDNA to the streptavidin coated PS beads (Bangs Laboratories). The reverse primers were unmodified. Grafting was obtained using a binding buffer (Dynabeads Kilobase binder Kit, Invitrogen). Biotin end-modified dsDNA fragments were mixed in a picomole range with the PS beads suspension in appropriate amounts to obtain a final volume of about 35μl and incubated at room temperature under gentle rotation for 12 hours in order to avoid sedimentation. The unreacted dsDNA fragments were removed using sequential washes with Milli-Q water. This can be easily achieved by centrifuging the suspension and by carefully pipetting off the supernatant and by finally resuspending the DNA coated beads to 40μl Milli-Q water. This procedure was repeated three times. The number of attached dsDNA chains per bead (functionality $f$) was quantified, knowing the number of the beads (value that can be determined by the concentration of the stock bead solution) and the number of DNA chains in the reaction vial before the cleaning procedure. The DNA concentration was determined by measuring the absorbance at 260 nm employing a micro-volume spectrometer (MicroDrop, ThermoScientific). Grafted particles were then dispersed in deionized water or a saline buffer solution containing $2.5 \times 10^{-2}$ M of NaCl. Dispersions with different particle concentrations were obtained by diluting a sediment obtained by centrifugation. The average area packing fraction $\eta$ of the confined dispersions was determined through the analysis of sample images by particle tracking. For each dispersion the results of the analysis of 1000 images were averaged.

**Microscopy experiments**

Quasi-2D samples were obtained by confining the dispersions between a microscope slide and a #1 coverslip: The distance between slide and coverslip was controlled by means of a PET-based double-sided tape with thickness h = 10 μm (No. 5601, Nitto). Glass surfaces were made hydrophobic through cleaning with Rain X solution (ITW Krafft) to avoid particle sticking to the glass. After depositing a 1.2 μl droplet of sample onto the microscope slide, the coverslip was uniformly pressed against the slide until reaching the desired separation and successively glued on the sides using epoxy resin. Microscopy experiments were performed on a Nikon Ti-S inverted microscope using a Nikon 50x LWD objective (N.A. 0.9). For each sample, about 50 series of 1000 images of 1280x1024 pixels were acquired at different locations in the sample using a 2.2Mp Pixelink M2 camera at a frame rate of 15fps. Particle coordinates were extracted from images using the Mosaic Suite for Fiji [65] while particle trajectories were determined using TrackMate [66]. De-drifting procedures available in TrackMate were applied to sample trajectories before calculating the MSDs. In order to avoid sample degradation, experiments were run shortly after sample loading.

**Monte-Carlo simulations**

Monte Carlo (MC) simulations employing the standard Metropolis algorithm were performed for soft discs in two dimensions interacting with the pair potential of Eq. (7), cut off at a distance $r_c = 3.5\sigma$, at constant temperature. The parameters of the potential determining the strength of the short-ranged attraction and the long-ranged repulsion as well as its range are reported in Fig. 5. The particles are contained in a box of dimensions $L_x = L_y = 20\sigma_0$. The number of particles is $N = \{196,$

289, 342, 380, 441, 600} in the systems with packing fraction η = {0.40, 0.57, 0.68, 0.73, 0.87, 1.17}, respectively. Data were gathered for simulation runs of $10^5$ MC steps for packing fractions η = 0.40 and 0.57; $10^6$ MC steps for η = 0.68, 0.73, and 0.87; and 2 x $10^6$ MC steps for the system with η = 1.17. Equilibration is achieved after 20%-50% of the given MC runs. The steric interaction diameter σ as a function of packing fraction is reported in the legend of Fig. 5A.

## ASSOCIATED CONTENT

**Supporting Information**.

Additional renderings of experimental samples (Fig.S1), additional radial distribution functions and mean squared displacements of experimental samples (Figs. S2-S3), additional theoretical details, additional information and results of molecular dynamics simulations (Figs. S4-S5) (one PDF file)

## AUTHOR INFORMATION

**Corresponding Authors**

*marco.laurati@unifi.it  *christos.likos@univie.ac.at.

**Author Contributions**

‡These authors contributed equally. I.R-S. prepared samples, performed experiments and analyzed the data. M.L., E.S. and L-E.C. instructed on sample preparation and experimental techniques, and provided materials. I.P. and C.N.L. developed the theoretical model of the PE-brush. I.P. performed the theoretical calculations and the single-brush MD simulations. N.A. performed the MC simulations of the effective model. All authors interpreted the results and

wrote the paper. M.L. and C.N.L. designed the research. The manuscript was written through contributions of all authors. All authors have given approval to the final version of the manuscript.

**Funding Sources**

Conacyt, "Consorzio per lo Sviluppo dei Sistemi a Grande Interfase" (CSGI), Austrian Science Fund (FWF), Deutsche Forschungsgemeinschaft (DFG).

**ACKNOWLEDGMENT**

We thank R. Castañeda-Priego, R. Moctezuma-Martiñon and P. van der Schoot for fruitful discussions. I.R-S., L-E.C. and M.L. acknowledge Conacyt for funding through grant A1-S-9098. M.L. and I.R-S. acknowledge financial support from "Consorzio per lo Sviluppo dei Sistemi a Grande Interfase" (CSGI). N.A. and C.N.L acknowledge support by the Austrian Science Fund (FWF) under grant I 2866-N36. E.S. acknowledges financial support by the Deutsche Forschungsgemeinschaft (DFG) under grant STI 664/4-1.

**REFERENCES**

(1) Ballauff, M.; Borisov, O. Polyelectrolyte Brushes. *Curr. Opin. Colloid Interface Sci.* **2006**, *11* (6), 316–323. https://doi.org/https://doi.org/10.1016/j.cocis.2006.12.002.

(2) Das, S.; Banik, M.; Chen, G.; Sinha, S.; Mukherjee, R. Polyelectrolyte Brushes: Theory, Modelling, Synthesis and Applications. *Soft Matter* **2015**, *11* (44), 8550–8583. https://doi.org/10.1039/C5SM01962A.

(3) Yang, W.; Zhou, F. Polymer Brushes for Antibiofouling and Lubrication. *Biosurface and Biotribology* **2017**, *3* , 97–114. https://doi.org/https://doi.org/10.1016/j.bsbt.2017.10.001.

(4) Wong, S. Y.; Han, L.; Timachova, K.; Veselinovic, J.; Hyder, M. N.; Ortiz, C.; Klibanov, A. M.; Hammond, P. T. Drastically Lowered Protein Adsorption on Microbicidal Hydrophobic/Hydrophilic Polyelectrolyte Multilayers. *Biomacromolecules* **2012**, *13* (3), 719–726. https://doi.org/10.1021/bm201637e.

(5) Krishnamoorthy, M.; Hakobyan, S.; Ramstedt, M.; Gautrot, J. E. Surface-Initiated Polymer Brushes in the Biomedical Field: Applications in Membrane Science, Biosensing, Cell Culture, Regenerative Medicine and Antibacterial Coatings. *Chem. Rev.* **2014**, *114* (21), 10976–11026. https://doi.org/10.1021/cr500252u.

(6) Wittemann, A.; Ballauff, M. Interaction of Proteins with Linear Polyelectrolytes and Spherical Polyelectrolyte Brushes in Aqueous Solution. *Phys. Chem. Chem. Phys.* **2006**, *8* (45), 5269–5275. https://doi.org/10.1039/B609879G.

(7) Wittemann, A.; Haupt, B.; Ballauff, M. Adsorption of Proteins on Spherical Polyelectrolyte Brushes in Aqueous Solution. *Phys. Chem. Chem. Phys.* **2003**, *5* (8), 1671–1677. https://doi.org/10.1039/B300607G.

(8) Lu, Y.; Ballauff, M. Spherical Polyelectrolyte Brushes as Nanoreactors for the Generation of Metallic and Oxidic Nanoparticles: Synthesis and Application in Catalysis. *Prog. Polym. Sci.* **2016**, *59*, 86–104. https://doi.org/https://doi.org/10.1016/j.progpolymsci.2016.03.002.

(9) Zhulina, E. B.; Rubinstein, M. Lubrication by Polyelectrolyte Brushes. *Macromolecules* **2014**, *47* (16), 5825–5838. https://doi.org/10.1021/ma500772a.

(10) Raviv, U.; Giasson, S.; Kampf, N.; Gohy, J.-F.; Jérôme, R.; Klein, J. Lubrication by Charged Polymers. *Nature* **2003**, *425* (6954), 163–165. https://doi.org/10.1038/nature01970.

(11) Yu, J.; Jackson, N. E.; Xu, X.; Morgenstern, Y.; Kaufman, Y.; Ruths, M.; de Pablo, J. J.; Tirrell, M. Multivalent Counterions Diminish the Lubricity of Polyelectrolyte Brushes. *Science* **2018**, *360* (6396), 1434–1438. https://doi.org/10.1126/science.aar5877.

(12) Ballauff, M. More Friction for Polyelectrolyte Brushes. *Science* **2018**, *360* (6396), 1399–1400. https://doi.org/10.1126/science.aat5343.

(13) Seror, J.; Zhu, L.; Goldberg, R.; Day, A. J.; Klein, J. Supramolecular Synergy in the Boundary Lubrication of Synovial Joints. *Nat. Commun.* **2015**, *6* (1), 6497. https://doi.org/10.1038/ncomms7497.

(14) Macdonald, M. L.; Samuel, R. E.; Shah, N. J.; Padera, R. F.; Beben, Y. M.; Hammond, P. T. Tissue Integration of Growth Factor-Eluting Layer-by-Layer Polyelectrolyte Multilayer Coated Implants. *Biomaterials* **2011**, *32* (5), 1446–1453. https://doi.org/10.1016/j.biomaterials.2010.10.052.

(15) Drummond, T. G.; Hill, M. G.; Barton, J. K. Electrochemical DNA Sensors. *Nat. Biotechnol.* **2003**, *21* (10), 1192–1199. https://doi.org/10.1038/nbt873.

(16) Maune, H. T.; Han, S.; Barish, R. D.; Bockrath, M.; III, W. A. G.; Rothemund, P. W. K.; Winfree, E. Self-Assembly of Carbon Nanotubes into Two-Dimensional Geometries Using DNA Origami Templates. *Nat. Nanotechnol.* **2010**, *5* (1), 61–66. https://doi.org/10.1038/nnano.2009.311.

(17) Tjong, V.; Tang, L.; Zauscher, S.; Chilkoti, A. "Smart" DNA Interfaces. *Chem. Soc. Rev.* **2014**, *43* (5), 1612–1626. https://doi.org/10.1039/C3CS60331H.

(18) Seeman, N. C.; Sleiman, H. F. DNA Nanotechnology. *Nat. Rev. Mater.* **2017**, *3* (1), 17068. https://doi.org/10.1038/natrevmats.2017.68.

(19) Rothemund, P. W. K. Folding DNA to Create Nanoscale Shapes and Patterns. *Nature* **2006**, *440* (7082), 297–302. https://doi.org/10.1038/nature04586.

(20) Chidchob, P.; Sleiman, H. F. Recent Advances in DNA Nanotechnology. *Curr. Opin. Chem. Biol.* **2018**, *46*, 63–70. https://doi.org/https://doi.org/10.1016/j.cbpa.2018.04.012.

(21) Nakata, M.; Zanchetta, G.; Chapman, B. D.; Jones, C. D.; Cross, J. O.; Pindak, R.; Bellini, T.; Clark, N. A. End-to-End Stacking and Liquid Crystal Condensation of 6– to 20–Base Pair DNA Duplexes. *Science* **2007**, *318* (5854), 1276–1279. https://doi.org/10.1126/science.1143826.

(22) Kilchherr, F.; Wachauf, C.; Pelz, B.; Rief, M.; Zacharias, M.; Dietz, H. Single-Molecule Dissection of Stacking Forces in DNA. *Science* **2016**, *353* (6304), aaf5508. https://doi.org/10.1126/science.aaf5508.

(23) Salamonczyk, M.; Zhang, J.; Portale, G.; Zhu, C.; Kentzinger, E.; Gleeson, J. T.; Jakli, A.; De Michele, C.; Dhont, J. K. G.; Sprunt, S.; Stiakakis, E. Smectic Phase in Suspensions of Gapped DNA Duplexes. *Nat. Commun.* **2016**, *7* (1), 13358. https://doi.org/10.1038/ncomms13358.

(24) Jones, M. R.; Mirkin, C. A. Self-Assembly Gets New Direction. *Nature* **2012**, *491* (7422),

42–43. https://doi.org/10.1038/491042a.

(25) Zhang, X.; Wang, R.; Xue, G. Programming Macro-Materials from DNA-Directed Self-Assembly. *Soft Matter* **2015**, *11* (10), 1862–1870. https://doi.org/10.1039/C4SM02649G.

(26) Rogers, W. B.; Shih, W. M.; Manoharan, V. N. Using DNA to Program the Self-Assembly of Colloidal Nanoparticles and Microparticles. *Nat. Rev. Mater.* **2016**, *1* (3), 16008. https://doi.org/10.1038/natrevmats.2016.8.

(27) Michele, L. Di; Eiser, E. Developments in Understanding and Controlling Self Assembly of DNA-Functionalized Colloids. *Phys. Chem. Chem. Phys.* **2013**, *15* (9), 3115–3129. https://doi.org/10.1039/C3CP43841D.

(28) Tan, S. J.; Kahn, J. S.; Derrien, T. L.; Campolongo, M. J.; Zhao, M.; Smilgies, D.-M.; Luo, D. Crystallization of DNA-Capped Gold Nanoparticles in High-Concentration, Divalent Salt Environments. *Angew. Chemie Int. Ed.* **2014**, *53* (5), 1316–1319. https://doi.org/https://doi.org/10.1002/anie.201307113.

(29) Pelaez-Fernandez, M.; Souslov, A.; Lyon, L. A.; Goldbart, P. M.; Fernandez-Nieves, A. Impact of Single-Particle Compressibility on the Fluid-Solid Phase Transition for Ionic Microgel Suspensions. *Phys. Rev. Lett.* **2015**, *114* (9), 98303. https://doi.org/10.1103/PhysRevLett.114.098303.

(30) Scotti, A.; Gasser, U.; Herman, E. S.; Pelaez-Fernandez, M.; Han, J.; Menzel, A.; Lyon, L. A.; Fernández-Nieves, A. The Role of Ions in the Self-Healing Behavior of Soft Particle Suspensions. *Proc. Natl. Acad. Sci.* **2016**, *113* (20), 5576–5581. https://doi.org/10.1073/pnas.1516011113.

(31) de With, A.; Greulich, K. O. Wavelength Dependence of Laser-Induced DNA Damage in Lymphocytes Observed by Single-Cell Gel Electrophoresis. *J. Photochem. Photobiol. B.* **1995**, *30* (1), 71–76. https://doi.org/10.1016/1011-1344(95)07151-q.

(32) Repula, A.; Oshima Menegon, M.; Wu, C.; van der Schoot, P.; Grelet, E. Directing Liquid Crystalline Self-Organization of Rodlike Particles through Tunable Attractive Single Tips. *Phys. Rev. Lett.* **2019**, *122* (12), 128008. https://doi.org/10.1103/PhysRevLett.122.128008.

(33) Zhang, J.; Lettinga, P. M.; Dhont, J. K. G.; Stiakakis, E. Direct Visualization of Conformation and Dense Packing of DNA-Based Soft Colloids. *Phys. Rev. Lett.* **2014**, *113* (26), 268303. https://doi.org/10.1103/PhysRevLett.113.268303.

(34) Moreno-Guerra, J. A.; Romero-Sánchez, I. C.; Martinez-Borquez, A.; Tassieri, M.; Stiakakis, E.; Laurati, M. Model-Free Rheo-AFM Probes the Viscoelasticity of Tunable DNA Soft Colloids. *Small* **2019**, *15* (42), 1904136. https://doi.org/https://doi.org/10.1002/smll.201904136.

(35) Likos, C. N. Effective Interactions in Soft Condensed Matter Physics. *Phys. Rep.* **2001**, *348* (4), 267–439. https://doi.org/https://doi.org/10.1016/S0370-1573(00)00141-1.

(36) Seth, J. R.; Mohan, L.; Locatelli-Champagne, C.; Cloitre, M.; Bonnecaze, R. T. A Micromechanical Model to Predict the Flow of Soft Particle Glasses. *Nat. Mater.* **2011**, *10* (11), 838–843. https://doi.org/10.1038/nmat3119.

(37) Bouhid de Aguiar, I.; van de Laar, T.; Meireles, M.; Bouchoux, A.; Sprakel, J.; Schroën, K. Deswelling and Deformation of Microgels in Concentrated Packings. *Sci. Rep.* **2017**, *7* (1), 10223. https://doi.org/10.1038/s41598-017-10788-y.

(38) Gasser, U.; Hyatt, J. S.; Lietor-Santos, J.-J.; Herman, E. S.; Lyon, L. A.; Fernandez-Nieves, A. Form Factor of PNIPAM Microgels in Overpacked States. *J. Chem. Phys.* **2014**, *141* (3), 34901. https://doi.org/10.1063/1.4885444.

(39) Conley, G. M.; Aebischer, P.; Nöjd, S.; Schurtenberger, P.; Scheffold, F. Jamming and Overpacking Fuzzy Microgels: Deformation, Interpenetration, and Compression. *Sci. Adv.* **2017**, *3* (10), e1700969. https://doi.org/10.1126/sciadv.1700969.

(40) Conley, G. M.; Zhang, C.; Aebischer, P.; Harden, J. L.; Scheffold, F. Relationship between Rheology and Structure of Interpenetrating, Deforming and Compressing Microgels. *Nat. Commun.* **2019**, *10* (1), 2436. https://doi.org/10.1038/s41467-019-10181-5.

(41) Nikolov, S. V; Fernandez-Nieves, A.; Alexeev, A. Behavior and Mechanics of Dense Microgel Suspensions. *Proc. Natl. Acad. Sci.* **2020**, *117* (44), 27096–27103. https://doi.org/10.1073/pnas.2008076117.

(42) Nöjd, S.; Holmqvist, P.; Boon, N.; Obiols-Rabasa, M.; Mohanty, P. S.; Schweins, R.; Schurtenberger, P. Deswelling Behaviour of Ionic Microgel Particles from Low to Ultra-High Densities. *Soft Matter* **2018**, *14* (20), 4150–4159. https://doi.org/10.1039/C8SM00390D.

(43) Manning, G. S. Limiting Laws and Counterion Condensation in Polyelectrolyte Solutions I. Colligative Properties. *J. Chem. Phys.* **1969**, *51* (3), 924–933. https://doi.org/10.1063/1.1672157.

(44) Rubinstein, M.; Colby, R. H. *Polymer Physics*; OUP Oxford, 2003.

(45) Jusufi, A.; Likos, C. N.; Ballauff, M. Counterion Distributions and Effective Interactions of Spherical Polyelectrolyte Brushes. *Colloid Polym. Sci.* **2004**, *282* (8), 910–917. https://doi.org/10.1007/s00396-004-1129-9.

(46) Wilk, A.; Huißmann, S.; Stiakakis, E.; Kohlbrecher, J.; Vlassopoulos, D.; Likos, C. N.; Meier, G.; Dhont, J. K. G.; Petekidis, G.; Vavrin, R. Osmotic Shrinkage in Star/Linear Polymer Mixtures. *Eur. Phys. J. E* **2010**, *32* (2), 127–134. https://doi.org/10.1140/epje/i2010-10607-2.

(47) Maffeo, C.; Luan, B.; Aksimentiev, A. End-to-End Attraction of Duplex DNA. *Nucleic Acids Res.* **2012**, *40* (9), 3812–3821. https://doi.org/10.1093/nar/gkr1220.

(48) De Michele, C. Theory of Self-Assembly-Driven Nematic Liquid Crystals Revised. *Liq. Cryst.* **2019**, *46* (13–14), 2003–2012. https://doi.org/10.1080/02678292.2019.1645366.

(49) Mossa, S.; Sciortino, F.; Tartaglia, P.; Zaccarelli, E. Ground-State Clusters for Short-Range Attractive and Long-Range Repulsive Potentials. *Langmuir* **2004**, *20* (24), 10756–10763. https://doi.org/10.1021/la048554t.

(50) Sciortino, F.; Mossa, S.; Zaccarelli, E.; Tartaglia, P. Equilibrium Cluster Phases and Low-Density Arrested Disordered States: The Role of Short-Range Attraction and Long-Range Repulsion. *Phys. Rev. Lett.* **2004**, *93* (5), 55701. https://doi.org/10.1103/PhysRevLett.93.055701.

(51) Vlassopoulos, D. Colloidal Star Polymers: Models for Studying Dynamically Arrested States in Soft Matter. *J. Polym. Sci. Part B Polym. Phys.* **2004**, *42* (16), 2931–2941. https://doi.org/https://doi.org/10.1002/polb.20152.

(52) Scheffold, F. Pathways and Challenges towards a Complete Characterization of Microgels. *Nat. Commun.* **2020**, *11* (1), 4315. https://doi.org/10.1038/s41467-020-17774-5.

(53) Ma, S.; Zhang, X.; Yu, B.; Zhou, F. Brushing up Functional Materials. *NPG Asia Mater.* **2019**, *11* (1), 24. https://doi.org/10.1038/s41427-019-0121-2.

(54) Liu, G.; Cai, M.; Zhou, F.; Liu, W. Charged Polymer Brushes-Grafted Hollow Silica Nanoparticles as a Novel Promising Material for Simultaneous Joint Lubrication and Treatment. *J. Phys. Chem. B* **2014**, *118* (18), 4920–4931. https://doi.org/10.1021/jp500074g.

(55) Nykypanchuk, D.; Maye, M. M.; van der Lelie, D.; Gang, O. DNA-Guided Crystallization of Colloidal Nanoparticles. *Nature* **2008**, *451* (7178), 549–552. https://doi.org/10.1038/nature06560.

(56) Macfarlane, R. J.; Lee, B.; Jones, M. R.; Harris, N.; Schatz, G. C.; Mirkin, C. A. Nanoparticle Superlattice Engineering with DNA. *Science* **2011**, *334* (6053), 204–208. https://doi.org/10.1126/science.1210493.

(57) Di Michele, L.; Varrato, F.; Kotar, J.; Nathan, S. H.; Foffi, G.; Eiser, E. Multistep Kinetic Self-Assembly of DNA-Coated Colloids. *Nat. Commun.* **2013**, *4* (1), 2007. https://doi.org/10.1038/ncomms3007.

(58) Woo, S.; Rothemund, P. W. K. Programmable Molecular Recognition Based on the Geometry of DNA Nanostructures. *Nat. Chem.* **2011**, *3* (8), 620–627. https://doi.org/10.1038/nchem.1070.

(59) Gerling, T.; Wagenbauer, K. F.; Neuner, A. M.; Dietz, H. Dynamic DNA Devices and Assemblies Formed by Shape-Complementary, Non–Base Pairing 3D Components. *Science* **2015**, *347* (6229), 1446–1452. https://doi.org/10.1126/science.aaa5372.

(60) Fan, J. A.; Wu, C.; Bao, K.; Bao, J.; Bardhan, R.; Halas, N. J.; Manoharan, V. N.; Nordlander, P.; Shvets, G.; Capasso, F. Self-Assembled Plasmonic Nanoparticle Clusters. *Science* **2010**, *328* (5982), 1135–1138. https://doi.org/10.1126/science.1187949.

(61) Kuzyk, A.; Schreiber, R.; Zhang, H.; Govorov, A. O.; Liedl, T.; Liu, N. Reconfigurable 3D Plasmonic Metamolecules. *Nat. Mater.* **2014**, *13* (9), 862–866. https://doi.org/10.1038/nmat4031.

(62) Sun, D.; Tian, Y.; Zhang, Y.; Xu, Z.; Sfeir, M. Y.; Cotlet, M.; Gang, O. Light-Harvesting Nanoparticle Core–Shell Clusters with Controllable Optical Output. *ACS Nano* **2015**, *9* (6), 5657–5665. https://doi.org/10.1021/nn507331z.

(63) Grzelczak, M.; Pérez-Juste, J.; Mulvaney, P.; Liz-Marzán, L. M. Shape Control in Gold Nanoparticle Synthesis. *Chem. Soc. Rev.* **2008**, *37* (9), 1783–1791. https://doi.org/10.1039/B711490G.

(64) Sacanna, S.; Korpics, M.; Rodriguez, K.; Colón-Meléndez, L.; Kim, S.-H.; Pine, D. J.; Yi, G.-R. Shaping Colloids for Self-Assembly. *Nat. Commun.* **2013**, *4* (1), 1688. https://doi.org/10.1038/ncomms2694.

(65) Sbalzarini, I. F.; Koumoutsakos, P. Feature Point Tracking and Trajectory Analysis for Video Imaging in Cell Biology. *J. Struct. Biol.* **2005**, *151* (2), 182–195. https://doi.org/https://doi.org/10.1016/j.jsb.2005.06.002.

(66) Tinevez, J.-Y.; Perry, N.; Schindelin, J.; Hoopes, G. M.; Reynolds, G. D.; Laplantine, E.; Bednarek, S. Y.; Shorte, S. L.; Eliceiri, K. W. TrackMate: An Open and Extensible Platform for Single-Particle Tracking. *Methods* **2017**, *115*, 80–90. https://doi.org/https://doi.org/10.1016/j.ymeth.2016.09.016.

**Supporting Information for**

**Blunt-end driven re-entrant ordering in quasi two-dimensional dispersions of spherical DNA brushes**

*Ivany Romero-Sanchez,[1,2]† Ilian Pihlajamaa,[3,4]† Natasa Adžić,[3] Laura Edith Castellano,[2] Emmanuel Stiakakis,[5] Christos N. Likos,[3]\* Marco Laurati[1]\**

[1]Dipartimento di Chimica & CSGI, Università di Firenze, 50019 Sesto Fiorentino, Italy

[2]División de Ciencias e Ingenierías, Universidad de Guanajuato, 37150 León, Mexico

[3]Faculty of Physics, University of Vienna, Bolzmanngasse 5, A-1090 Vienna, Austria

[4]Eindhoven University of Technology, Department of Applied Physics, Soft Matter and Biological Physics, Postbus 513, NL-5600 MB Eindhoven, The Netherlands

[5]Biomacromolecular Systems and Processes, Institute of Biological Information Processing (IBI-4), 4 Forschungszentrum Jülich, D-52425 Jülich, Germany

**This PDF file includes:**

Supplementary text
Figures S1 to S5
SI References

**Supporting Information Text**

**Additional exemplary renderings of experimental samples**

We report in Fig. S1 exemplary renderings of the experimental sample structures complementing those reported in Fig.3 of the main article. Renderings were reconstructed from the coordinates extracted from particle tracking applied to optical microscopy experiments.

**Radial distribution functions $g(r)$ and mean squared displacements for the system with 25mM NaCl**

We show in Fig. S2 the $g(r)$ functions obtained from the particle coordinates extracted from microscopy experiments on samples containing 25mM NaCl in solution. Compared to the system without salt, the samples show a gradual and monotonic increase of the order with increasing $\eta$, as indicated by the progressive increase of the height of the first peak and the shift of the peak position to increasingly lower values.
The corresponding mean squared displacements are presented in Fig. S3. Similar to the $g(r)$, the MSDs curves for different value of $\eta$ show a progressive, almost monotonic (except for sample with $\eta = 0.70$) reduction of the displacements.

**Derivation of the free energy model**

As mentioned in the main text, we include six contributions in the free energy

$$F = U_\text{H} + F_\text{el} + F_\text{Fl} + S_2 + S_3 + F_\text{p}.$$

Here, we elaborate on the electrostatic Hartree term $U_\text{H}$ and the entropic terms $S_2$ and $S_3$. Additionally, we provide the analytical expression that results from the minimization procedure with respect to the number of escaped counterions $N_3$ and the brush height $L$.

The Hartree energy is given by

$$U_\text{H} = \frac{1}{8\pi\varepsilon} \iint \text{d}r \, \text{d}r' \frac{\rho(\text{r})\rho(\text{r}')}{|r-r'|},$$

where

$$\rho(r) = \begin{cases} 0 & r < R_{PS} \\ -\dfrac{eN_3}{4\pi L r^2} & R_{PS} \leq r \leq R \\ \dfrac{3eN_3}{4\pi(R_W^3 - R^3)} & R < r < R_W \end{cases}$$

as explained in the main text. The Hartree energy can now be evaluated directly by substitution of $\rho(r)$. The result is

$$\frac{U_\text{H}}{k_B T} = \frac{N_3^2 \lambda_B}{2} \left[ \frac{R_{PS}}{L^2} \vartheta_1\left(\frac{R}{R_{PS}}\right) + \frac{1}{R} \vartheta_2\left(\frac{R}{R_W}\right) \right]$$

in which the functions $\vartheta_1(x)$ and $\vartheta_2(x)$ are given by

$$\vartheta_1(x) = x - 2\ln x - \frac{1}{x}, \quad \text{and} \quad \vartheta_2(x) = \frac{5 - 9x + 5x^3 - x^6}{5(1 - x^3)^2}.$$

Since $L = R - R_{PS}$, the derivatives of these functions will appear in the result of the free energy minimization.

The entropic contribution in the free energy can be described as the sum of two contributions. The first, $S_2$, is the contribution due to the counterions within the brush that are not strongly condensed along the PE-chains. The second, $S_3$, is the term that corresponds to the counterions that have escaped the brush. Here, we neglect the entropy of the condensed counterions. However, we do take into account that the presence of the PE-chains limits the available free volume to the entropic counterions inside the brush. This can be achieved by introducing a reduced brush thickness $L'$, such that the available volume to the entropically active counterions is reduced by the volume of the PE-chains, which we model as cylindrical.

$$V_2 \equiv \frac{4\pi}{3}(R_{PS} + L')^3 = \frac{4\pi}{4}(R_{PS} + L)^3 - fL\pi R_m^2,$$

where $R_\text{m}$ is the radius of a chain monomer and $V_2$ is the volume available to the entropic counterions. The volume available to the counterions outside the brush is simply

$$V_3 = \frac{4\pi}{3}(R_W^3 - R^3).$$

The entropic contributions can now be evaluated from Eq. (9) in the main text

$$\frac{S_2}{k_B T} = N_2 \left[ 1 + \ln\left(\frac{N_2 \sigma^3}{4\pi L'(R_{PS} + L')^2}\right) - \frac{2R_{PS}}{L'} \ln\left(\frac{R_{PS} + L'}{R_{PS}}\right) \right]$$

$$\frac{S_3}{k_B T} = N_3 \left[ \ln\left(\frac{N_3 \sigma^3}{V_3}\right) - 1 \right].$$

As all contributions to the free energy in our model are now known, we can proceed with the minimization with respect to $N_3$ and $L$, under the condition that $N_1 + N_2 + N_3 = Nf$. This ultimately yields the system of equations

$$N_3 \lambda_B = \frac{2 + \ln\left(\frac{N_2}{N_3}\frac{R_W^3 - R^3}{3L'(R_{PS}+L')^2}\right) - \frac{2R_{PS}}{L'}\ln\left(\frac{R_{PS}+L'}{R_{PS}}\right) + \frac{R^3}{R_W^3}}{\frac{R_c}{L^2}\vartheta_1\left(\frac{R}{R_{PS}}\right) + \frac{1}{R}\vartheta_2\left(\frac{R}{R_W}\right)},$$

$$\frac{N_3^2 \lambda_B}{2} = \frac{\frac{3fL}{Nb^2} - \frac{9vf^2 N^2}{8\pi L^4} + \frac{2R_{PS}N_2}{L'^2}\ln\left(\frac{R_{PS}+L'}{R_{PS}}\right) - \frac{3N_2}{L'} + \frac{4\pi R^2 N_3}{V_3} + \frac{3N_2 R^2}{R_W^3}}{\frac{R_{PS}}{L^3}\vartheta_1\left(\frac{R}{R_{PS}}\right) - \frac{1}{L^2}\vartheta_1'\left(\frac{R}{R_{PS}}\right) + \frac{1}{R^2}\vartheta_2\left(\frac{R}{R_W}\right) - \frac{1}{RR_W}\vartheta_2'\left(\frac{R}{R_W}\right)},$$

which we solve numerically using Newton's method.

**Additional information and results on the molecular dynamics simulations of the miniature brush**

To verify the theoretical predictions that we make in the main text, we perform molecular dynamics simulations of miniature brushes. Our simulation model consists of a large colloidal particle to which $f$ charged polymer chains are grafted. The PE-chains are simulated similarly to the model that was used by Wynveen *et al.* (*65*)

We coarse-grain the PE-chains as a single polymer chain consisting of Lennard-Jones monomers that each have an assigned charge $-e$. To ensure charge neutrality, we introduce $fN$ positively charged counterions into the simulation box. All particles interact according to a shifted and truncated Lennard-Jones potential as function of the distance $r$ between the centers of mass of the particles. This potential models the steric interactions of particles in good solvent conditions, and is given by

$$V_{LJ}^{\mu\nu}(r) = \begin{cases} 4\varepsilon_{LJ}\left[\left(\frac{d}{r-\delta^{\mu\nu}}\right)^{12} - \left(\frac{d}{r-\delta^{\mu\nu}}\right)^6\right] + \varepsilon_{LJ} & \text{if} \quad r < \sqrt[6]{2}d + \delta^{\mu\nu} \\ 0 & \text{otherwise} \end{cases}$$

where we set $d = 4.0$ Å, and $\varepsilon_{LJ} = 1.0$ kJ/mol. The indices $\mu$ and $\nu$ represent the three different species of particles $\mu, \nu \in \{C_0, M_-, m_+\}$, in which $C_0$ denotes the large central colloid, $M_-$ the chain monomers, and $m_+$ the free counterions. The shift $\delta^{\mu\nu}$ in the Lennard-Jones potential (not to be confused with a Kronecker delta) is given by $\delta^{\mu\nu} = r^\mu + r^\nu - d$, where $r^{C_0} = R_{PS} = 100$ Å, $r^{M_-} = R_m = 9.0$ Å, and $r^{m_+} = d/2 = 2.0$ Å. This interaction potential models counterion-counterion interactions with a usual truncated Lennard-Jones potential. All other two-particle interactions are modelled with this same potential, shifted by $\delta^{\mu\nu}$. This choice implies that the monomer-counterion interaction is equal to $\varepsilon_{LJ}$ at $r = 11$ Å, consistent with the steric radius of a DNA-chain.

We model electrostatic interactions between the particles $i$ and $j$ having charge numbers $Z_i$ and $Z_j$ with a Coulomb potential

$$V_C^{ij}(r) = k_B T Z_i Z_j \frac{\lambda_B}{r},$$

in which we treat the solvent implicitly by a dielectric medium at temperature $k_B T = 2.4\varepsilon_{LJ}$ with corresponding Bjerrum length $\lambda_B = 7.1$ Å, characteristic of water at room temperature. The long-range interactions were computed using a three-dimensional Ewald-summation with a relative error smaller than $10^{-4}$. Decreasing this value is found to have no significant effect on the simulation results. Periodic boundaries are implemented using the minimum image convention.

The monomers in the chains are connected to their neighbors by harmonic bonds that are described by the harmonic potential

$$V_b(r) = \frac{1}{2} k_H (r - b)^2,$$

where $b = 3.4$ Å and $k_H = 210 \,\text{kJ mol}^{-1}\text{Å}^{-2}$. The latter value was chosen such that potential has a strength of $k_B T$ when $|r - b| = 0.15$ Å, which is a measure of the dispersion of the monomers of a chain. Because the equilibrium bond length is much smaller than the distance at which the steric monomer-monomer interaction diverges, we exclude the repulsive interactions of the 8 nearest neighbors on each side of every chain particle. To be able to capture the swelling behavior of the brush due to the osmotic pressure of the counterions and the electrostatic repulsion between the chain monomers, we do not include a valence angle potential.

Using a Langevin thermostat, we effectively perform Brownian dynamics simulations in the canonical ensemble. The time step is set to $10^{-3}$, where $\tau = \sqrt{md^2/\varepsilon_{LJ}} \approx 10$ ps, with $m = 660$ g/mol as the mass of a chain monomer, roughly equal to the mass of a DNA base pair. The masses of the colloid and counterions respectively are set to $6.6 \times 10^4$ g/mol and 20 g/mol. We set the relaxation time of the Langevin equations to equal $\tau$, which implicitly defines the effective viscosity of the solvent. The simulations were typically run for $\sim 10^6$ time steps, excluding an equilibration time of $10^5 - 10^6$ steps.

In the construction of the theoretical model, a few important assumptions were made. One of those stated that the density of monomers and counterions within the brush decays as $r^{-2}$, where $r$ is the distance from the brush center. Our simulation results confirm this as is shown in Fig S4. Here we see the monomer and counterion number densities plotted on a double logarithmic scale, and we find that, within the brush, the monomer density decays roughly as $r^{-2.0}$. Note that the exponent of $-2$ does not strictly mean that the chains are perfectly stretched; a brush with chains that have 'corkscrew' features, or stretched chains that are not perfectly radially oriented also produce the same exponent in the decay of the monomer density. Both phenomena are to some degree present in the simulation results, as is confirmed by snapshots of the simulation, of which an example is shown in Fig. 2b in the main text.

In Fig. S4, we also observe that the counterion distribution follows the profile of the monomer distribution within the brush to minimize the electrostatic interaction energy. In contradiction to our assumptions, the counterion distribution does not appear to be constant outside of the brush. This has however a negligible effect on the long-range electrostatic interaction and on the entropy $S_3$. The aforementioned model has been also applied to two interacting miniature brushes to calculate in a molecular dynamics simulation the effective force between them as a function of their separation. Results are shown in Fig. S5 in comparison with the theoretical result, confirming the validity of the latter. Due to its small number of arms, the miniature brush is not osmotic, and thus the interaction is dominated in this case by electrostatics. For the experimental osmotic brushes, on the other hand, the potential is a steeply increasing function of the separation as the latter diminishes beyond brush overlap, as established earlier by Jusufi *et al.* (2)

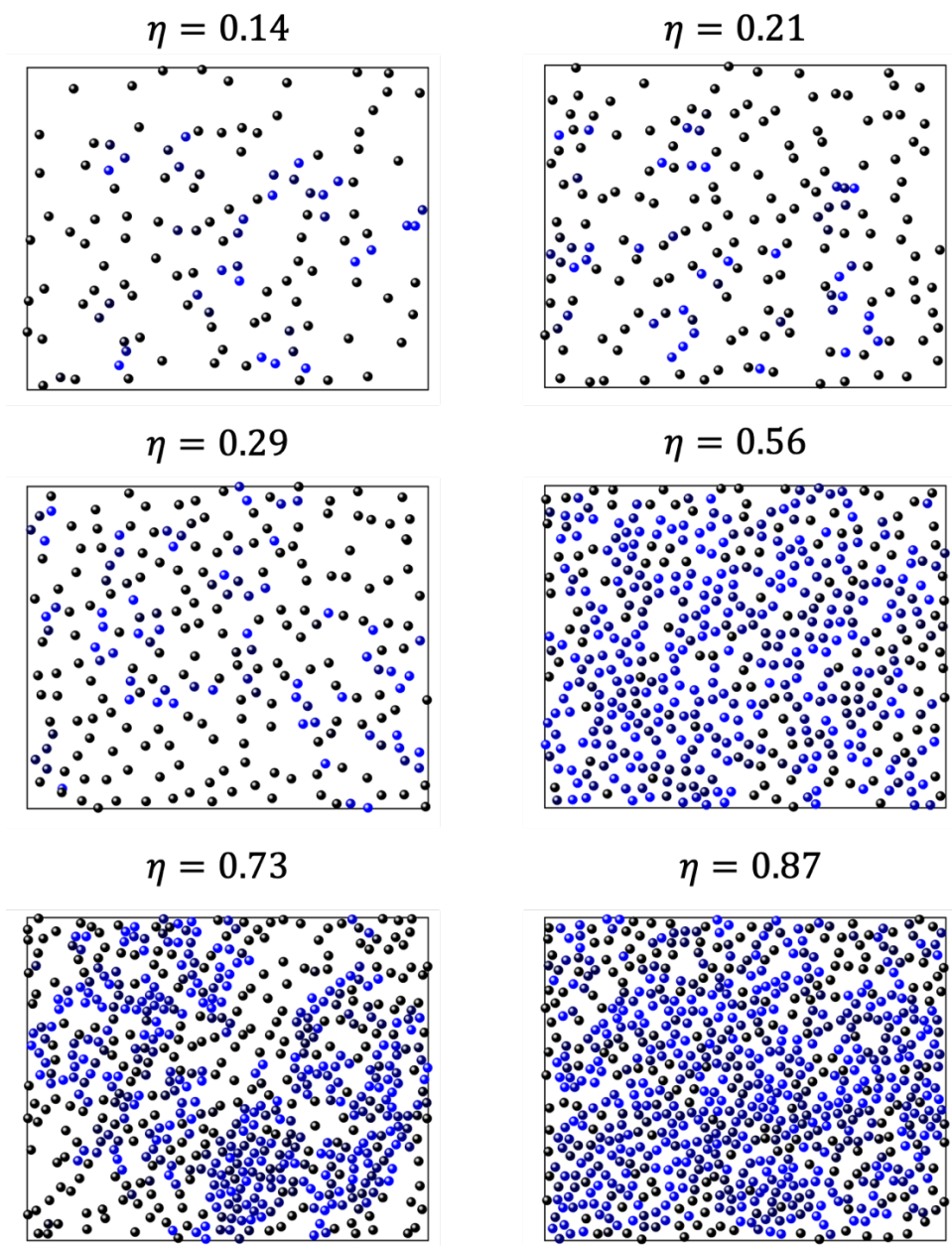

**Fig. S1.** Renderings of the experimental samples, representative of the structural organization of the dispersion with the corresponding packing fraction $\eta$, as indicated.

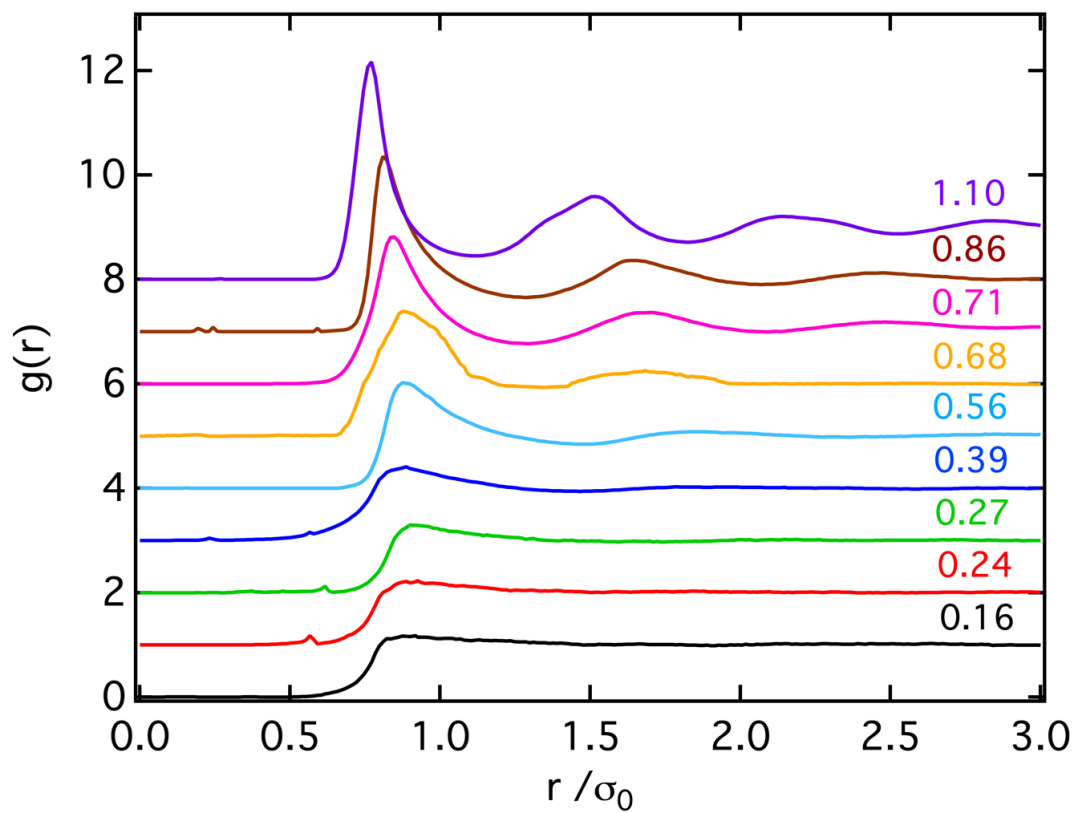

**Fig. S2.** Radial distribution functions $g(r)$ for the system with 25mM NaCl, for different packing fractions $\eta$, as indicated. The curves are shifted each time by a factor of 1 with respect to the data corresponding to the previous value of $\eta$.

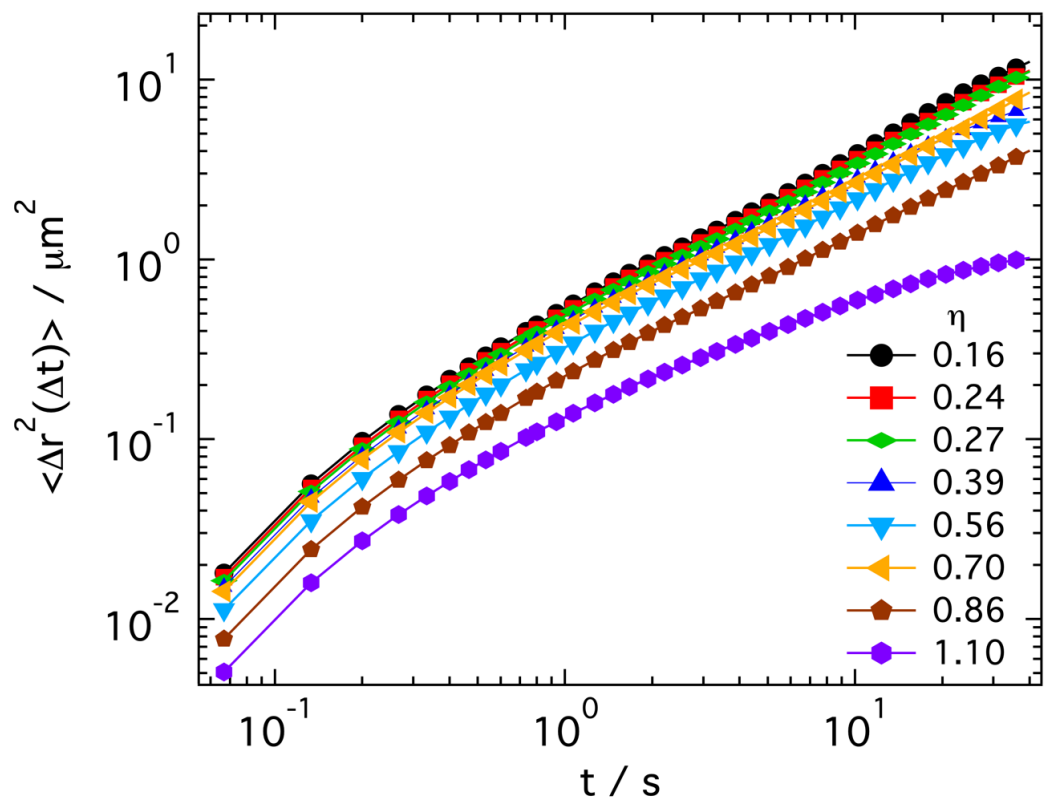

**Fig. S3.** Mean squared displacements $\langle \Delta r^2(\Delta t) \rangle$ for dispersions of dsDNA colloids with 25mM NaCl in solution. The packing fraction is indicated in the legend.

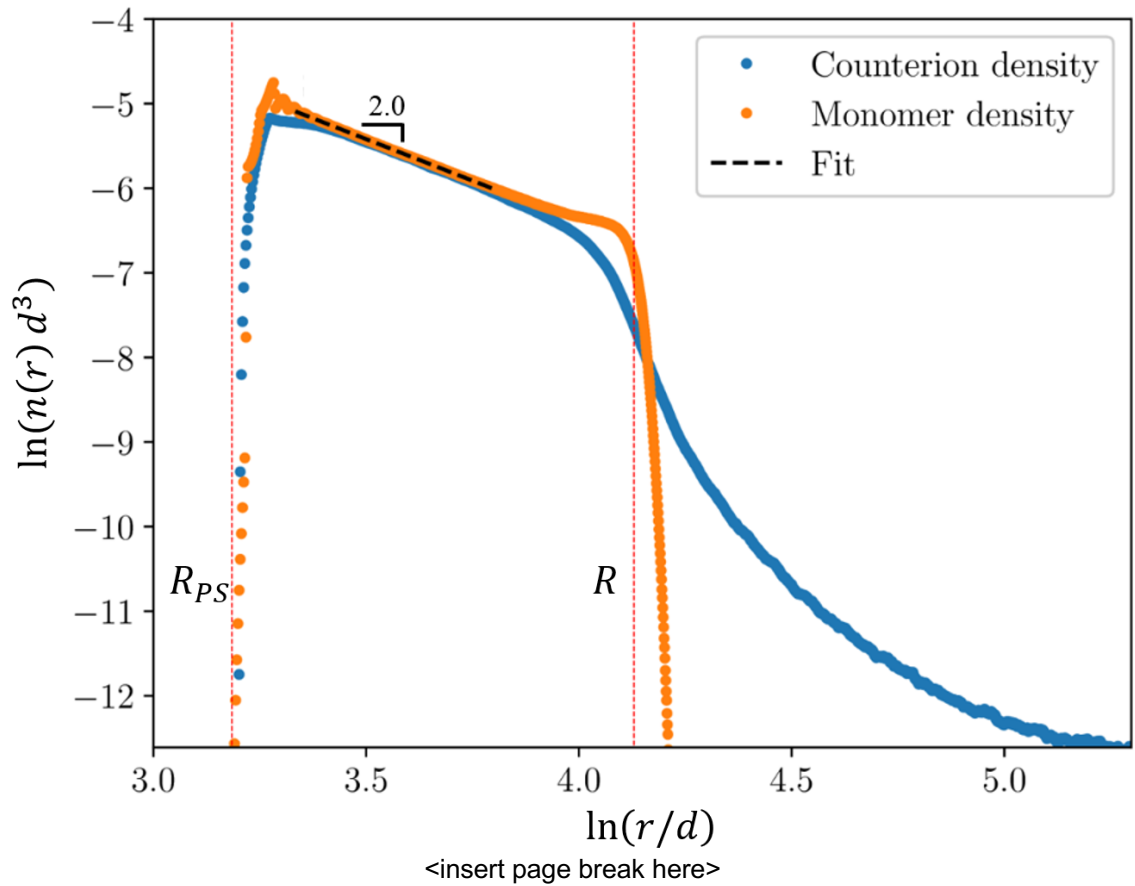

**Fig. S4.** Double logarithmic plot of the monomer and counterion number density as function of the radial distance for a brush with functionality $f = 40$ and chain length $N = 60$ in a cubic box with sides $L_x = L_y = L_z = 400$. The vertical dashed lines indicate the colloid radius $R_{PS}$ and the brush radius $R = L + R_{PS}$, the latter being determined as the average distance the last chain monomers are separated from the brush centre. A straight line was fitted to the monomer density data within the brush. We find a slope of -2.0, indicating the stretched nature of the PE chains.

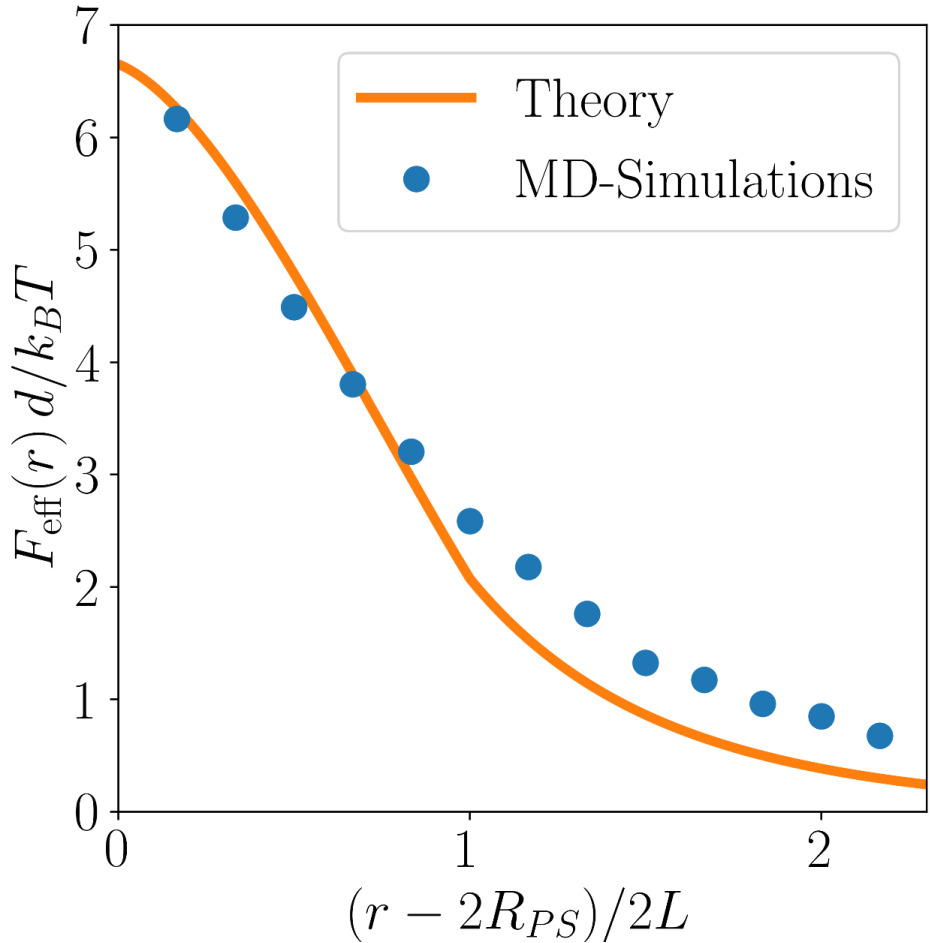

**Fig. S5.** The effective force $F_{\text{eff}}$ in units of $k_B T/d$ as a function of the separation distance $r$ acting on brushes with functionality $f = 20$ and chain length $N = 50$. We have fixed the box size at $L_x/4 = L_y = L_z = 160d$. The dots are molecular dynamics simulation results, whereas the line represents the predictions from the theory (derivation will be provided upon reasonable request). The horizontal axis is scaled such that a value of 0 indicates that the two colloidal cores are touching, $r = 2R_{PS}$. Similarly, a value of 1 indicates that the brushes are touching but not overlapping, $r = 2R$.

**References**

1. A. Wynveen, C. N. Likos, Interactions between planar polyelectrolyte brushes: effects of stiffness and salt. *Soft Matter* **6**, 163–171 (2010).
2. A. Jusufi, C. N. Likos, M. Ballauff, Counterion distributions and effective interactions of spherical polyelectrolyte brushes. *Colloid Polym. Sci.* **282**, 910–917 (2004).